# Homogenization of periodic hexa- and tetrachiral cellular solids

Andrea Bacigalupo[1], Luigi Gambarotta[2*]

[1]*Department of Civil, Environmental and Mechanical Engineering, University of Trento*
*Via Mesiano, 77, 38123, Trento, Italy*

[2]*Department of Civil, Chemical and Environmental Engineering, University of Genoa*
*Via Montallegro, 1, 16145, Genoa, Italy*

**Abstract**

The homogenization of auxetic cellular solids having periodic hexachiral and tetrachiral microstructure is dealt with two different techniques. The first approach is based on the representation of the cellular solid as a beam-lattice to be homogenized as a micropolar continuum. The second approach is developed to analyse periodic cells conceived as a two-dimensional domain consisting of deformable portions such as the ring, the ligaments and possibly an embedded matrix internally to these. This approach is based on a second displacement gradient computational homogenization proposed by the Authors (Bacigalupo and Gambarotta, 2010). The elastic moduli obtained by the micropolar homogenization are expressed in analytical form from which it appears explicitly their dependence on the parameter of chirality, which is the angle of inclination of the ligaments with respect to the grid of lines connecting the centers of the rings. For hexachiral cells, the solution of Liu *et al.*, 2012, is found, showing the auxetic property of the lattice together with the elastic coupling modulus between the normal and the asymmetric strains; a property that has been confirmed here for the tetrachiral lattice. Unlike the hexagonal lattice, the classical constitutive equations of the tetragonal lattice turns out to be characterized by the coupling between the normal and shear strains through an elastic modulus that is an odd function of the parameter of chirality. Moreover, this lattice is found to exhibit a remarkable variability of the Young's modulus and of the Poisson's ratio with the direction of the applied uniaxial stress. The properties of the equivalent micropolar continuum are qualitatively detected also in the equivalent second-gradient continuum. Moreover, for both the hexachiral and the tetrachiral cellular material, the second-order elastic moduli obtained through the homogenization technique are in agreement with the invariance properties defined by Auffray *et al.*, 2009. This investigation, that is justified by the need of understanding the effects of the compliance of the rings and of the filling material, has shown that it is sufficient a very soft filling material to get significant increases in the Poisson's ratio, until to lose the auxetic property of these cellular solids. Finally, the experimental and numerical results by Alderson *et al.*, 2010, are compared to the theoretical ones obtained by the homogenization techniques here considered.

* Corresponding author: Luigi Gambarotta, luigi.gambarotta@unige.it

# 1. Introduction

Auxetic materials, having zero or negative Poisson's ratio, are characterized by non-conventional mechanical response with respect to many common materials: they become thicker widthwise when stretched lengthwise and thinner when compressed (Prawoto, 2012). Although some natural materials may be classified as auxetic, this quality is mostly obtained in man-made materials (Greaves *et al.*, 2011). This unusual mechanical behaviour may result in an increased stiffness and indentation resistance of the auxetic material and a higher toughness due to an increase of the energy absorption under static and dynamic loading, thus making these smart materials of special technological interest (see Scarpa *et al.*, 2013). The auxetic effect occurs in cellular materials, such as foams (see Lakes, 1987), honeycomb structures and networks (Smith *et al.*, 2000) and origami structures (Schenk and Guest, 2013), as the result of the unfolding of re-entrant cells as they are stretched. The design of auxetic materials is mostly addressed to periodic cellular composites (Cadman *et al.*, 2013) through the analysis and optimization of periodic manufacturable cells (Andreassen *et al.*, 2014, Xu *et al.*, 1999). In addition to the periodic microstructures based on re-entrant mechanisms, auxetic materials based on mechanisms of rotating rigid and semi-rigid units (Grima *et al.*, 2005) and on rolling-up mechanisms (Prall and Lakes, 1997) have been developed. This latter mechanism occurs in two-dimensional honeycomb structures composed of circular rings periodically located in the material plane and joined by straight ligaments characterized by chiral (see figure 1) or anti-chiral topologies.

Alderson *et al.*, 2010, carried out experiments on samples having both chiral and anti-chiral periodic cells subjected to uniaxial compression, together with numerical simulations of the experimental results obtained by a standard FE homogenization of the periodic cell. While a rather good agreement in the overall elastic moduli was found for the hexachiral cell (Figure 1.a), qualitative differences were obtained between the experimental and numerical results for the tetrachiral cell. Further theoretical and experimental analyses have been carried out by Lorato *et al.* , 2010, and Cicala *et al.*, 2012, to investigate the transverse elastic properties of chiral honeycombs, and by Chen *et al.*, 2013, to derive the in-plane elastic moduli of anti-tetrachiral lattices.

With reference to the chiral topologies, the study of the mechanical behavior of hexachiral structures started from the seminal paper by Prall and Lakes, 1997, and was later developed to include the analysis of damage processes (see Bettini *et al.*, 2010) and of free wave propagation (Spadoni *et al.*, 2009). Afterwards, Tee *et al.*, 2010, to obtain the phononic properties of the tetrachiral periodic cell, applied a FE analysis based on the Floquet-Bloch approach. An analysis of the overall elastic properties of chiral and anti-chiral cellular solids was carried out by Dirrenberger *et al.*, 2011, 2013, that is based on the classical homogenisation approach through a finite element discretization of the periodic cell. This approach has been extended by Dirrenberger *et al.*, 2012, to the analysis of the elasto-plastic response of hexachiral ductile materials.

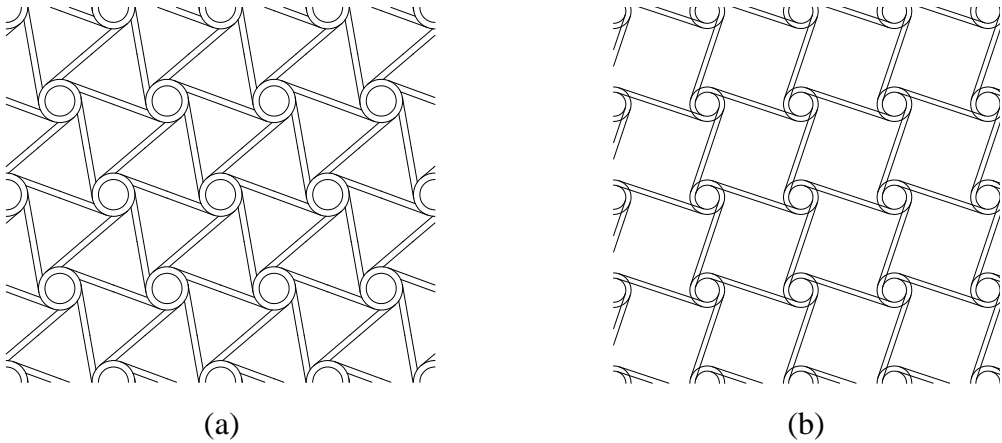

(a)            (b)

Figure 1. (a) Hexachiral lattice; (b) tetrachiral lattice.

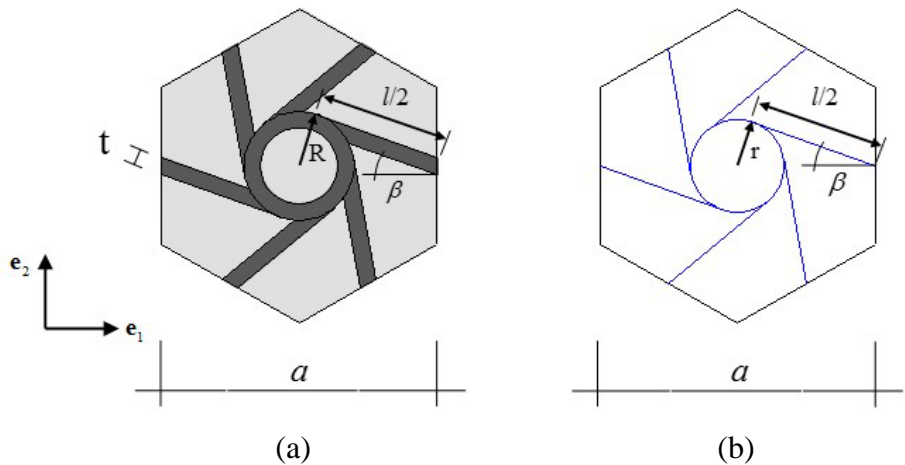

(a)            (b)

Figure 2. Hexachiral periodic cells: (a) two-dimensional; (b) beam-lattice.

Spadoni and Ruzzene, 2012, developed a micropolar homogenization of the hexachiral beam-lattice model (see figure 2.b) based on the approach of Kumar and McDowell, 2004, that is equipped with an internal length directly associated with the characteristic size of the microstructure. The overall elastic moduli of the equivalent micropolar continuum were found to depend on the chirality parameter $\beta$ measured by the angle of inclination of the ligaments with respect to the grid of lines connecting the centers of the rings. Moreover, the classical elastic moduli of the equivalent transversely isotropic continuum, i.e., the overall Young's modulus and Poisson's ratio, were derived so improving the estimation of the Poisson's ratio obtained from Prall and Lakes, 1997. A further improvement of the micropolar homogenization of hexachiral beam-lattice was obtained by Liu *et al.*, 2012, which have shown that the chiral geometry determines a coupling between the bulk deformation and the pure rotation. This effect is described by an elastic modulus that is an odd function of the chirality angle $\beta$, namely it reverses its sign when the material pattern is flipped over. Despite this improvement of the micropolar model, the resulting overall elastic moduli of the classical continuum are unchanged from those obtained by Spadoni and Ruzzene, 2012.

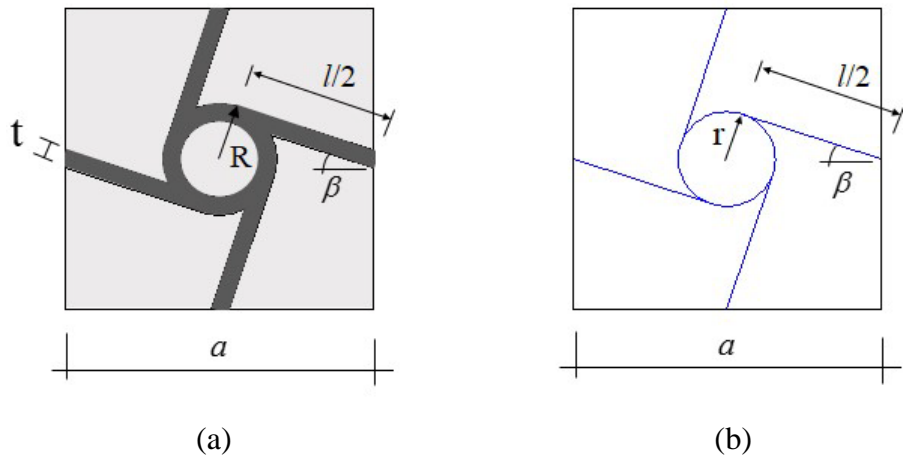

(a)            (b)

Figure 3. Tetrachiral periodic cells: (a) two-dimensional; (b) beam-lattice.

The results by Liu *et al.*, 2012, and Spadoni and Ruzzene, 2012, concerning the hexachiral beam-lattice model (figure 2b) provide a richer description of the dependence of the elastic moduli on the chirality and deserve to be extended to the tetrachiral geometry (Figure 2b). On the other hand, the beam-lattice model can be regarded as appropriate for very slender ligaments, a circumstance that does not seem to occur in some samples used in experiments where the effective length of the ligament is not easy to identify. Furthermore, this model does not include the presence of the filling material between the ligaments and inside of the rings. For these reasons it seems necessary to define an equivalent continuum at the macroscale, preferably a non-local continuum equipped with internal lengths, which is based on a FE description of the periodic cell. Although a smart technique for micropolar homogenization of two-dimensional cells (Forest and Sab, 1998) is available, there are considerations that limit its use (see Trinh *et al.*, 2012) and suggest computational homogenization techniques based on second gradient continuum models (see Bacigalupo and Gambarotta, 2010, Bacca *et al.*, 2013, Bacigalupo and Gambarotta, 2013, Bacigalupo, 2014). To support this choice, some studies aimed to define the elastic properties of chiral materials according to the strain gradient elasticity (Papanicolopulos, 2011, Auffray *et al.*, 2009, 2010, 2013) may be regarded as a reference for the validation of numerical simulations.

In this paper the overall elastic moduli for both hexachiral and thetrachiral periodic cells (figure 2, 3) are obtained with reference to both the micropolar and second displacement gradient continuum models. At first, the cellular materials are modeled as beam lattices having rigid circular rings and elastic beams with rigid ends to represent the ligaments and a micropolar equivalent continuum is obtained. Afterwards, the problem of computational homogenization of chiral cells with thick ligaments and filled with a soft matrix is addressed through a second-gradient homogenization technique proposed by the Authors (Bacigalupo and Gambarotta, 2010). Here, generalized boundary conditions of periodicity are introduced which guarantee the continuity of the micro-displacement field at the interface of adjacent cells. For both the hexachiral and the tetrachiral cellular solids it is shown the dependence of the elastic moduli on the chirality parameter $\beta$. Moreover, a comparison of the elastic moduli provided by the micro-polar and second gradient approaches is given with reference to the case of symmetric stresses, i.e. to the elastic moduli of the classical continuum. A further comparison concerns the different auxetic behavior of the two considered cellular arrays: the auxetic isotropy of the hexachiral model and the auxetic orthotropy of the tethrachiral model. Finally, the experimental results by Alderson *et al.*, 2010, are compared to the theoretical ones obtained by the homogenization techniques considered in this paper and those by Alderson *et al.*, 2010.

## 2. Micropolar modelling of periodic chiral cellular solids

The chiral lattices shown in Figure 1 may be considered as beam-lattices modeled as a two-dimensional micro-polar continuum. This description requires the ligaments to be sufficiently slender and is formulated on the hypothesis of rigid rings, as developed by Spadoni and Ruzzene, 2012, and Liu et al., 2012, for the case of hexachiral lattice. The $n$ ligaments of the periodic cell, shown in Figure 4 with reference to the hexachiral material $n = 6$, are modeled as elastic beams rigidly connected to the rings. The ligament is tangent to the joined rigid rings with length $l$ between the connection points, section width $t$ and thickness $d$, Young's modulus $E_s$. More specifically, the ligament is modeled as a beam having rigid ends with length e to represent the portion of the ligament connected to the ring where a limited deformation occurs. The central portion of the ligament is assumed to be elastic having length $l_0 = l - 2e = \rho l$, where the ratio $\rho = \dfrac{l_0}{l}$ is introduced.

The displacements at the ends of the *i*-th ligament that connects the central reference ring with the *i*-th adjacent ring are expressed in the form (see figure 4)

$$\mathbf{u}_i^0 = \mathbf{U} - \phi\, r\, \mathbf{d}_i \quad , \quad \mathbf{u}_i = \mathbf{U}_i + \phi_i\, r\, \mathbf{d}_i \tag{1}$$

and involve the displacements of the rigid circular rings, i.e. the displacement $\mathbf{U}$ and the rotation $\phi$ of the central ring and the displacement $\mathbf{U}_i$ and rotation $\phi_i$ of the *i*-th ring, respectively, the unit vector $\mathbf{d}_i$ associated with the ligament and the radius $r$ of the ring. The displacement vector and the rotation of the *i*-th ring are approximated by affine fields of macro-displacement and rotation, namely based on a first-order expansion

$$\begin{cases} \mathbf{U}_i = \mathbf{U} + \mathbf{H}\mathbf{x}_i \\ \phi_i = \phi + \mathbf{k}\bullet\mathbf{x}_i \end{cases}, \tag{2}$$

$\mathbf{H} = \nabla \mathbf{U}$ being the macro-displacement gradient and $\mathbf{k} = \nabla \phi$ the curvature, respectively, playing the role of macro-strain variables (see Kumar and McDowell, 2004). Consequently, the relative displacement between the ends of the ligament depends on the rotation $\phi$ and on the gradients $\mathbf{H}$ and $\mathbf{k}$ in the form

$$\mathbf{u}_i - \mathbf{u}_i^0 = \mathbf{H}\mathbf{x}_i + r\left(\mathbf{d}_i \otimes \mathbf{x}_i\right)\mathbf{k} + 2\phi\, r\, \mathbf{d}_i \ . \qquad (3)$$

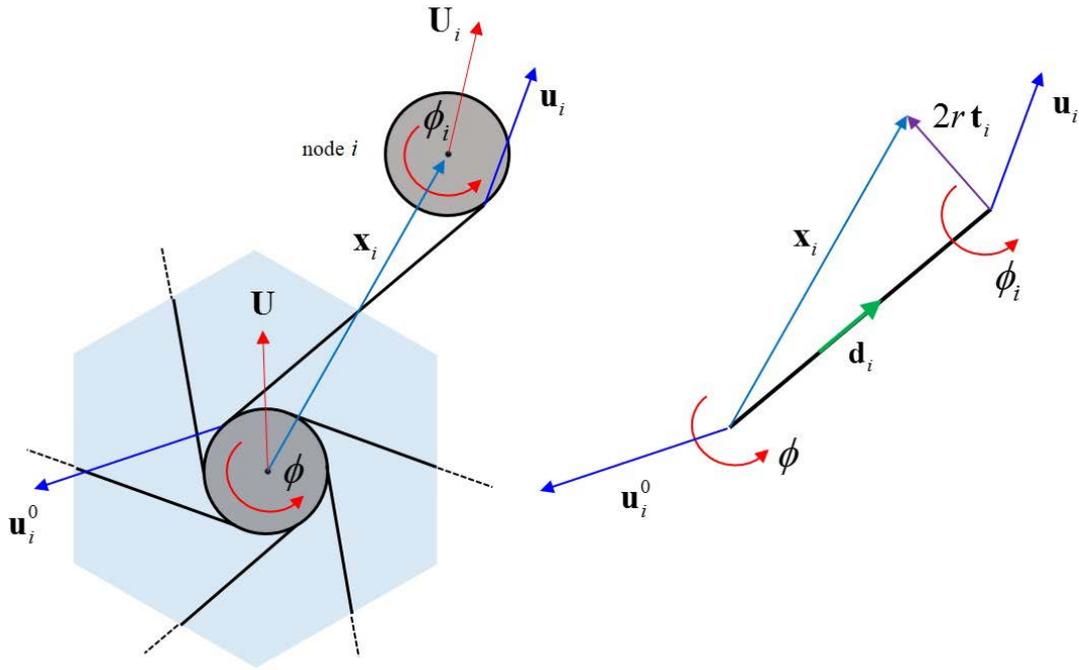

Figure 4. The *i*-th ligament between two adjacent cells: cell dofs and beam end displacements.

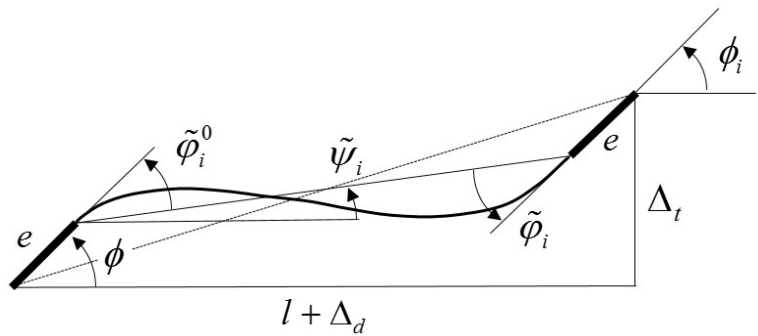

Figure 5. Elastic ligament with rigid offsets and generalized displacements.

The ligament extension is
$\Delta_{di} = (\mathbf{u}_i - \mathbf{u}_i^0) \cdot \mathbf{d}_i = (\mathbf{d}_i \otimes \mathbf{x}_i) : \mathbf{\Gamma} + r\mathbf{x}_i \cdot \mathbf{k} = a(\mathbf{d}_i \otimes \mathbf{n}_i) : \mathbf{\Gamma} + ar\mathbf{n}_i \cdot \mathbf{k}$, where $\mathbf{\Gamma} = \mathbf{H} - \mathbf{W}(\phi)$ is introduced as the Cosserat strain tensor, $\mathbf{W}$ the rotation tensor (with non-vanishing components $w_{12} = -\phi$ and $w_{21} = \phi$), $\mathbf{n}_i$ the unit vector related to the line connecting the centres of two adjacent cells having length $a$, i.e. $\mathbf{x}_i = a\,\mathbf{n}_i$. The transverse relative displacement between the ends of the ligament is $\Delta_{ti} = (\mathbf{u}_i - \mathbf{u}_i^0) \cdot \mathbf{t}_i = \mathbf{t}_i \cdot \mathbf{H}\mathbf{x}_i = a\,\mathbf{t}_i \cdot \mathbf{H}\mathbf{n}_i$, being $\mathbf{t}_i = \frac{1}{2r}\mathbf{x}_i - \frac{1}{2r}\mathbf{d}_i$ ($\mathbf{t}_i \cdot \mathbf{d}_i = 0$). The axial strain of the deformable portion of the ligament shown in figure 5 is $\varepsilon_i = \frac{\Delta_{di}}{l_0} = \frac{\Delta_{di}}{\rho l}$. The mean rotation of the deformable portion is written as $\tilde{\psi}_i = (\Delta_{ti} - \phi e - \phi_i e)/l_0$ (see figure 5) and the ends' rotations are

$$\tilde{\varphi}_i^0 = \phi - \tilde{\psi}_i = \phi - \frac{1}{\rho}\psi_i + \frac{1-\rho}{2\rho}(\phi_i + \phi) =$$
$$= -\frac{1}{\rho l}(\mathbf{t}_i \otimes \mathbf{x}_i) : \mathbf{\Gamma} + \frac{1-\rho}{2\rho}\mathbf{x}_i \cdot \mathbf{k} = -\frac{a}{\rho l}(\mathbf{t}_i \otimes \mathbf{n}_i) : \mathbf{\Gamma} + \frac{1-\rho}{2\rho}a\,\mathbf{n}_i \cdot \mathbf{k},$$
$$\tilde{\varphi}_i = \phi_i - \tilde{\psi}_i = \phi_i - \frac{1}{\rho}\psi_i + \frac{1-\rho}{2\rho}(\phi_i + \phi) = \quad (4)$$
$$= -\frac{1}{\rho l}(\mathbf{t}_i \otimes \mathbf{x}_i) : \mathbf{\Gamma} + \frac{1+\rho}{2\rho}\mathbf{x}_i \cdot \mathbf{k} = -\frac{a}{\rho l}(\mathbf{t}_i \otimes \mathbf{n}_i) : \mathbf{\Gamma} + \frac{1+\rho}{2\rho}a\,\mathbf{n}_i \cdot \mathbf{k},$$

with $\phi = \frac{1}{l}\mathbf{t}_i \cdot \mathbf{W}(\phi)\mathbf{x}_i$. The axial strain energy of the ligament takes the form $\mathcal{E}_{ai} = \frac{1}{2}E_s t d \varepsilon_i^2 l_0 = \frac{1}{2}\frac{E_s}{\rho}d\left(\frac{t}{l}\right)\Delta_{di}^2$, while the bending strain energy is $\mathcal{E}_{bi} = \frac{E_s t^2 d}{6\rho}\left(\frac{t}{l}\right)(\tilde{\varphi}_i^{0\,2} + \tilde{\varphi}_i^2 + \tilde{\varphi}_i^0\,\tilde{\varphi}_i)$. It follows that the total strain energy stored in the $i$-th ligament $\mathcal{E}_i = \mathcal{E}_{ai} + \mathcal{E}_{bi} = \frac{1}{2}\left(\mathbf{\Gamma}\bullet\tilde{\mathbb{C}}_i\mathbf{\Gamma} + \mathbf{k}\cdot\tilde{\mathbb{Y}}_i\mathbf{\Gamma} + \mathbf{\Gamma}\bullet\tilde{\mathbb{Y}}_i^T\mathbf{k} + \mathbf{k}\cdot\tilde{\mathbf{S}}_i\mathbf{k}\right)$ is a quadratic form in the Cosserat macrostrain tensors $\mathbf{\Gamma}$ and $\mathbf{k}$, that involves the fourth-order tensor $\tilde{\mathbb{C}}_i$, the fifth-order tensor $\tilde{\mathbb{Y}}_i$ and the sixth-order tensor of the ligament, respectively.

Hence, the total strain energy stored in the periodic cell is obtained by superimposing the contributions of the $n$ ligaments $\mathcal{E}_m = \sum_{i=1}^{n} \frac{1}{2} \mathcal{E}_i$.

By applying the macro-homogeneity criterion by Hill-Mandel, i.e. by equating the strain energy in the micropolar equivalent continuum $\mathcal{E}_M = \frac{1}{2} A \left( \boldsymbol{\Gamma} \bullet \tilde{\mathbb{C}} \boldsymbol{\Gamma} + \boldsymbol{\Gamma} \bullet \tilde{\mathbb{Y}} \mathbf{k} + \mathbf{k} \cdot \tilde{\mathbb{Y}}^T \boldsymbol{\Gamma} + \mathbf{k} \cdot \tilde{\mathbf{S}} \mathbf{k} \right)$, $A$ being the area of the periodic cell, to the total strain energy in the periodic cell at the microscale, namely $\mathcal{E}_M = \mathcal{E}_m$, the overall elasticity tensors for the micropolar equivalent continuum are obtained $\tilde{\mathbb{C}} = \sum_{i=1}^{n} \frac{1}{2} \tilde{\mathbb{C}}_i$, $\tilde{\mathbb{Y}} = \sum_{i=1}^{n} \frac{1}{2} \tilde{\mathbb{Y}}_i$, $\tilde{\mathbf{S}} = \sum_{i=1}^{n} \frac{1}{2} \tilde{\mathbf{S}}_i$. As known, in case of centrosymmetric cell the coupling fifth-order tensor turns out to be vanishing. Moreover, by the definition of strain energy density at the macroscale, the stress $\tilde{\boldsymbol{\Sigma}} = \partial \mathcal{E}_M / \partial \boldsymbol{\Gamma} = \tilde{\mathbb{C}} \boldsymbol{\Gamma} + \tilde{\mathbb{Y}} \mathbf{k}$ and the couple-stress $\mathbf{m} = \partial \mathcal{E}_M / \partial \mathbf{k} = \tilde{\mathbb{Y}}^T \boldsymbol{\Gamma} + \mathbf{S} \mathbf{k}$ tensors are derived. Finally, the constitutive equation of the micropolar continuum may be written according to the Voigt notation

$$\left\{ \begin{array}{c} \underline{\Sigma} \\ \underline{m} \end{array} \right\} = \left[ \begin{array}{cc} \underline{\tilde{C}} & \underline{\tilde{Y}} \\ \underline{\tilde{Y}}^T & \underline{\tilde{S}} \end{array} \right] \left\{ \begin{array}{c} \underline{\Gamma} \\ \underline{\chi} \end{array} \right\} \quad , \tag{5}$$

involving the macro-strain vectors $\underline{\Gamma} = \{ \Gamma_{11} \quad \Gamma_{22} \quad \Gamma_{12} \quad \Gamma_{21} \}^T$ and $\underline{k} = \{ \chi_1 \quad \chi_2 \}^T$, being $\Gamma_{11} = u_{1,1}$, $\Gamma_{22} = u_{2,2}$, $\Gamma_{12} = u_{1,2} + \phi$, $\Gamma_{21} = u_{2,1} - \phi$, $\chi_1 = \phi_{,1}$ e $\chi_2 = \phi_{,2}$, the macro-stress tensors $\underline{\tilde{\Sigma}} = \{ \tilde{\Sigma}_{11} \quad \tilde{\Sigma}_{22} \quad \tilde{\Sigma}_{12} \quad \tilde{\Sigma}_{21} \}^T$ and $\underline{m} = \{ m_1 \quad m_2 \}^T$ and the overall elasticity sub-matrices $\underline{\tilde{C}}$, $\underline{\tilde{Y}}$ and $\underline{\tilde{S}}$.

## 2.1 Hexachiral honeycomb

The constitutive equation of hexachiral honeycomb corresponds to that obtained by Liu *et al.*, 2012,

$$\begin{Bmatrix} \tilde{\Sigma}_{11} \\ \tilde{\Sigma}_{22} \\ \tilde{\Sigma}_{12} \\ \tilde{\Sigma}_{21} \\ m_1 \\ m_2 \end{Bmatrix} = \begin{bmatrix} 2\mu+\lambda & \lambda & -A & A & 0 & 0 \\ \lambda & 2\mu+\lambda & -A & A & 0 & 0 \\ -A & -A & \mu+\kappa & \mu-\kappa & 0 & 0 \\ A & A & \mu-\kappa & \mu+\kappa & 0 & 0 \\ 0 & 0 & 0 & 0 & S & 0 \\ 0 & 0 & 0 & 0 & 0 & S \end{bmatrix} \begin{Bmatrix} \Gamma_{11} \\ \Gamma_{22} \\ \Gamma_{12} \\ \Gamma_{21} \\ \chi_1 \\ \chi_2 \end{Bmatrix}, \quad (6)$$

in which the elastic moduli

$$\mu = \frac{\sqrt{3}}{4} E_s \delta \frac{(\rho^2+\delta^2)}{\rho^3},$$

$$\lambda = \frac{\sqrt{3}}{4} E_s \delta \frac{(\rho^2-\delta^2)}{\rho^3} (\cos^2\beta - \sin^2\beta),$$

$$\kappa = \frac{\sqrt{3}}{2} E_s \delta \frac{(\rho^2 \sin^2\beta + \delta^2 \cos^2\beta)}{\rho^3}, \quad (7)$$

$$S = \frac{\sqrt{3}}{12} E_s \delta a^2 \frac{\left[3\rho^2 \sin^2\beta + (3+\rho^2)\delta^2 \cos^2\beta\right]}{\rho^3},$$

$$A = -\frac{\sqrt{3}}{2} E_s \delta \frac{(\rho^2-\delta^2)}{\rho^3} \sin\beta \cos\beta,$$

depend here not only on the slenderness ratio $\delta = t/l$ and on the angle $\beta$ of inclination of the ligaments, but also on the ratio $\rho$ (this parameter is indirectly determined by the thickness $t$ of the ligament and by the radius $r$ of the ring). The constitutive equations (6) show the coupling between the extensional strains $\Gamma_{11}$ and $\Gamma_{22}$ and the asymmetric strains $\Gamma_{12}$ and $\Gamma_{21}$ through the elastic constant $A$, not considered in the formulation of Spadoni and Ruzzene, 2012. As already observed by Liu *et al.*, 2012, the elastic moduli, with the exception of $\mu$, depend on the parameter of chirality $\beta$, but only the constant $A$ is an odd function of this parameter, i.e. it reverses its sign when the handedness of the material pattern is flipped over. Moreover, it is easy to see that for $\beta = 0$ and $\rho = 1$ the results by Kumar and McDowell, 2004, for hexagonal honeycomb with $S>0$ are obtained (see also Bažant and Christensen, 1972, for the discussion concerning the second order stiffness parameter $S$).

In case of symmetric macro-strain fields $\mathbf{\Gamma} = \mathbf{\Gamma}^T$, with $\mathbf{W}(\phi) = skw\mathbf{H}$, the constitutive equation is the classical one $\mathbf{\Sigma} = \mathbb{C}\mathbf{E}$ referred to the Cauchy continuum, with $\mathbf{E} = sym\mathbf{\Gamma}$ and $\mathbf{\Sigma} = sym\tilde{\mathbf{\Sigma}}$. The fourth order elastic tensor for the hexagonal system corresponds to that of the transversely isotropic system whose elastic moduli in the plane of the lattice are:

$$E_{\text{hom}} = \frac{2\sqrt{3}E_s\delta^3\left(\rho^2 + \delta^2\right)}{\left(3\rho^2\delta^2 + \delta^4\cos^2\beta + \rho^4\sin^2\beta\right)\rho} ,$$

$$\nu_{\text{hom}} = \frac{\left(\delta^2\cos^2\beta - \rho^2\sin^2\beta\right)\left(\rho^2 - \delta^2\right)}{3\rho^2\delta^2 + \delta^4\cos^2\beta + \rho^4\sin^2\beta} , \qquad (8)$$

$$G_{\text{hom}} = \frac{\sqrt{3}}{4}E_s\delta\frac{\left(\rho^2 + \delta^2\right)}{\rho^3} .$$

This result, which is similar to that of Liu *et al.* (2012) and Spadoni and Ruzzene (2012), the latter obtained by a different way, shows that the overall Young's modulus and Poisson's ratio of the hexachiral honeycomb are even functions of the parameter of chirality $\beta$, thus these moduli are not able to distinguish the lefthanded from the righthanded lattice. Conversely, the shear modulus G is independent of the parameter of chirality, as already highlighted by Spadoni and Ruzzene, 2012, and Liu *et al.*, 2012, namely the hexachiral lattice does not exhibit an increase of shearing stiffness when the Poisson's ratio takes negative values. Furthermore, it should be noted that these results improve those derived by Prall and Lakes, 1997, on the basis of a classical Cauchy homogenization where the axial deformability of the ligaments was neglected.

The effect of the slenderness of the ligaments and of the angle of chirality $\beta$ on the overall moduli is summarized in the diagrams in Figures 6,7, and 8. In the diagrams of figure 6 it is shown the overall elastic modulus to increase with the slenderness ratio $\delta = t/l$ and to decrease with the modulus of chirality $|\beta|$ and with the parameter $\rho$. In addition, the diagrams in Figure 7 show that increasing the modulus of chirality $|\beta|$, the Poisson's ratio decreases up to get the auxetic behavior of the lattice. Finally, the diagram in Figure 8 shows that the elasticity shear modulus increases with the thickness of the ligament, i.e. with the ratio $\delta$, and decreases with the ratio $\rho$.

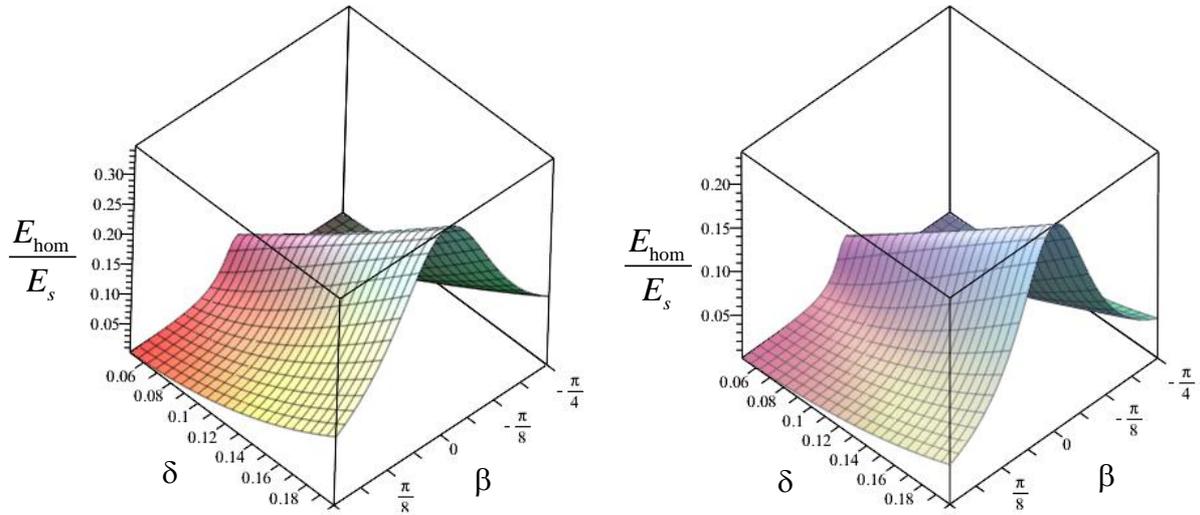

Figure 6. Hexachiral material: ratio between the overall in-plane Young modulus to the ligament Young modulus for varying chirality $\beta$ and ligament slenderness $\delta$; effect of the rigid offset: (a) $\rho = 0.7$, (b) $\rho = 1$.

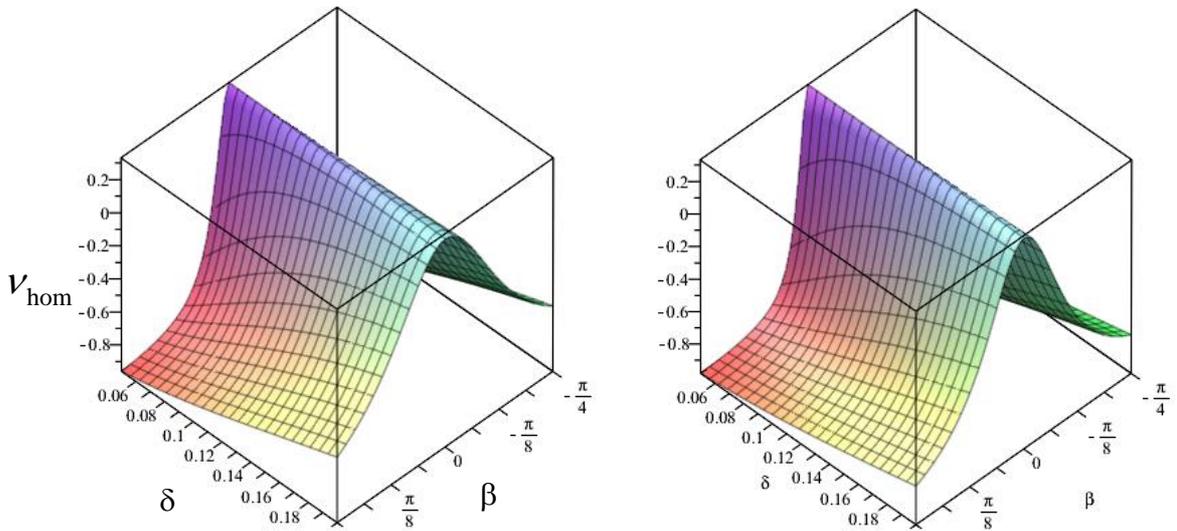

Figure 7. Hexachiral material: overall Poisson ratio for varying chirality $\beta$ and ligament slenderness $\delta$; effect of the rigid offset: (a) $\rho = 0.7$, (b) $\rho = 1$.

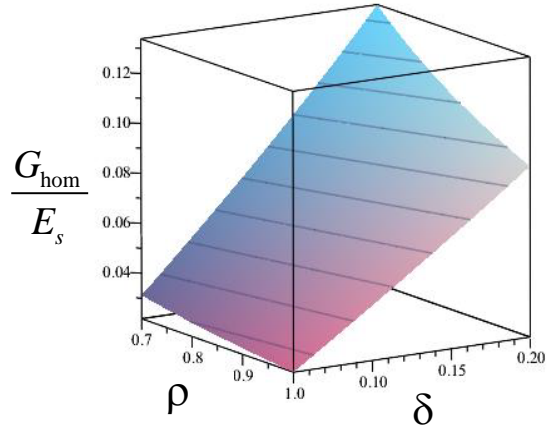

Figure 8. Hexachiral material: ratio between the overall in-plane shear modulus to the ligament Young modulus for varying the ligament slenderness $\delta$ and the extension of the rigid offset.

## 2.2 Tetrachiral honeycomb

The micropolar constitutive equation of the tetrachiral honeycomb shown in Figure 3.b takes the following form:

$$\begin{Bmatrix} \tilde{\Sigma}_{11} \\ \tilde{\Sigma}_{22} \\ \tilde{\Sigma}_{12} \\ \tilde{\Sigma}_{21} \\ m_1 \\ m_2 \end{Bmatrix} = \begin{bmatrix} 2\mu & 0 & 0 & B & 0 & 0 \\ 0 & 2\mu & -B & 0 & 0 & 0 \\ 0 & -B & \kappa & 0 & 0 & 0 \\ B & 0 & 0 & \kappa & 0 & 0 \\ 0 & 0 & 0 & 0 & S & 0 \\ 0 & 0 & 0 & 0 & 0 & S \end{bmatrix} \begin{Bmatrix} \Gamma_{11} \\ \Gamma_{22} \\ \Gamma_{12} \\ \Gamma_{21} \\ \chi_1 \\ \chi_2 \end{Bmatrix}, \qquad (9)$$

with

$$\mu = \frac{E_s \delta}{2\rho^3}\left(\rho^2 \cos^2 \beta + \delta^2 \sin^2 \beta\right),$$

$$\kappa = \frac{E_s \delta}{\rho^3}\left(\rho^2 \sin^2 \beta + \delta^2 \cos^2 \beta\right),$$

$$B = -\frac{E_s \delta}{\rho^3}\left(\rho^2 - \delta^2\right)\sin \beta \cos \beta,$$

$$S = \frac{1}{12}\frac{E_s \delta a^2}{\rho^3}\left[3\rho^2 \sin^2 \beta + \left(3+\rho^2\right)\delta^2 \cos^2 \beta\right]. \qquad (10)$$

Similarly to the hexachiral honeycomb, a coupling is obtained between the extensional strains and the asymmetric strains through the elastic modulus B which is an odd function of the parameter of chirality $\beta$, while the other elastic moduli are even functions.

In case of symmetric macro-strain fields $\boldsymbol{\Gamma} = \boldsymbol{\Gamma}^T$, the resulting classical fourth order elasticity tensor $\mathbb{C}$ has the elasticities of the tetragonal system generated by rotations $\mathbf{R}_3^{\frac{\pi}{2}}$ of $\pi/2$ around the unit vector $\mathbf{e}_3$ (Gurtin, 1972). In the Voigt notation, the constitutive equation takes the form

$$\begin{Bmatrix} \Sigma_{11} \\ \Sigma_{22} \\ \Sigma_{12}^s \end{Bmatrix} = \begin{bmatrix} C_{1111} & C_{1122} & C_{1112} \\ C_{1122} & C_{1111} & -C_{1112} \\ C_{1112} & -C_{1112} & C_{1212} \end{bmatrix} \begin{Bmatrix} E_{11} \\ E_{22} \\ 2E_{12} \end{Bmatrix}, \qquad (11)$$

with

$$C_{1111} = \frac{E_s \delta}{2\rho^3} \frac{\left(\rho^4 + \delta^4\right)\sin^2\beta\cos^2\beta + 2\rho^2\delta^2\left(\sin^2\beta + \cos^4\beta\right)}{\rho^2\sin^2\beta + \delta^2\cos^2\beta},$$

$$C_{1122} = -\frac{E_s \delta}{2\rho^3} \frac{\left(\rho^2 - \delta^2\right)^2 \sin^2\beta\cos^2\beta}{\rho^2\sin^2\beta + \delta^2\cos^2\beta}, \qquad (12)$$

$$C_{1212} = \frac{E_s \delta}{2\rho^3}\left(\rho^2\sin^2\beta + \delta^2\cos^2\beta\right),$$

$$C_{2212} = -C_{1112} = \frac{E_s \delta}{2\rho^3}\left(\rho^2 - \delta^2\right)\sin\beta\cos\beta.$$

The elasticity tensor depends on the chirality parameter, but unlike the hexachiral honeycomb, some elasticities are odd functions of $\beta$, namely $C_{2212} = -C_{1112}$. The fourth-order tensor of elastic compliance $\mathbb{D} = \mathbb{C}^{-1}$ depends on the parameter of chirality as well, but its non-vanishing components in the considered reference

$$D_{1111} = D_{2222} = \frac{\rho}{E_s \delta^3}\left(\rho^2\sin^2\beta + \delta^2\cos^2\beta\right),$$

$$D_{1212} = \frac{\rho}{E_s \delta^3}\left(\rho^2\cos^2\beta + \delta^2\sin^2\beta\right), \qquad (13)$$

$$D_{1112} = -D_{2212} = \frac{\rho\left(\rho^2 - \delta^2\right)}{2E_s \delta^3}\sin\beta\cos\beta.$$

show that the tetrachiral honeycomb exhibits an auxetic limit behaviour $\nu_{12} = \nu_{21} = 0$ being $D_{1122} = 0$. While the extension along the direction of the applied uniaxial tension is coupled with the asymmetric deformation, and thus with the rings rotation, conversely no transverse extension takes place ($\nu = -\varepsilon_{ii}/\varepsilon_{jj} = 0$, with $i \neq j$). A result that qualitatively differs from the ones by Alderson *et al.*, 2010, where a negative Poisson's ratio was obtained ($\nu \cong -0.35$).

A more complete description of the elastic response may be obtained by evaluating the effects of the uniaxial tension $\boldsymbol{\sigma} = \sigma \mathbf{n} \otimes \mathbf{n}$ applied along direction $\mathbf{n}$ identified by the angle $\theta$ with respect to the unit vector $\mathbf{e}_1$. Being $E_n(\theta) = \sigma(\mathbf{n} \otimes \mathbf{n}) : \mathbb{D}(\mathbf{n} \otimes \mathbf{n})$ the extension along $\mathbf{n}$ and $E_t(\theta) = \sigma(\mathbf{t} \otimes \mathbf{t}) : \mathbb{D}(\mathbf{n} \otimes \mathbf{n})$ the transverse extension, the overall elastic modulus along direction $\mathbf{n}$ is defined as

$$E_{\text{hom}}(\theta) = \sigma/E_n(\theta) = \left[(\mathbf{n} \otimes \mathbf{n}) : \mathbb{D}(\mathbf{n} \otimes \mathbf{n})\right]^{-1} \qquad (14)$$

together with the corresponding overall Poisson's ratio

$$\nu_{\text{hom}}(\theta) = -E_t(\theta)/E_n(\theta) = -\frac{(\mathbf{t} \otimes \mathbf{t}) : \mathbb{D}(\mathbf{n} \otimes \mathbf{n})}{(\mathbf{n} \otimes \mathbf{n}) : \mathbb{D}(\mathbf{n} \otimes \mathbf{n})} \quad . \qquad (15)$$

Both the ratio $E_{\text{hom}}(\theta)/E_s$ between the overall elastic modulus and the elastic modulus of the ligaments and the overall Poisson's ratio $\nu_{\text{hom}}(\theta)$ are shown in the non-dimensional diagrams of figures 9 and 10 for values of the material parameters related to those corresponding to the thetragonal honeycomb experimented by Alderson *et al.*, 2010. From the polar and Cartesian diagrams of figure 9.a and 9.b, respectively, the ratio $E_{\text{hom}}(\theta)/E_s$ markedly depends on the angle $\theta$. This indicates a strong anisotropy with well-defined directions of maximum stiffness which depend on the angle of chirality $\beta$. This parameter also affects the coefficient of transverse contraction as is shown in the diagrams of Figure 10. In particular, when the uniaxial tension is applied along the direction of minimum elastic stiffness, the Poisson's ratio is positive and attains the maximum value $\nu_{\text{hom}} \simeq 0.75$, and the material does not exhibit an auxetic behavior. Conversely, for the directions of the applied uniaxial tension corresponding to the maximum stiffness, the Poisson's ratio is negative and attains the minimum value $\nu_{hom} \simeq -0.9$. As already noted, for the particular case of uniaxial stress applied along the straight lines joining the centres of the rigid rings, the Poisson's ratio is vanishing $\nu_{hom} = 0$. Finally, the effect of the tetrachiral geometry of the lattice is a rotation of the directions of maximum stiffness and auxetic behaviour ($\nu_{hom}(\theta) < 0$) with respect to the grid of orthogonal axes formed by the straight lines joining the centres of the rings. As expected, for vanishing chirality $\beta = 0$, the directions of maximum stiffness correspond with the axes of this grid.

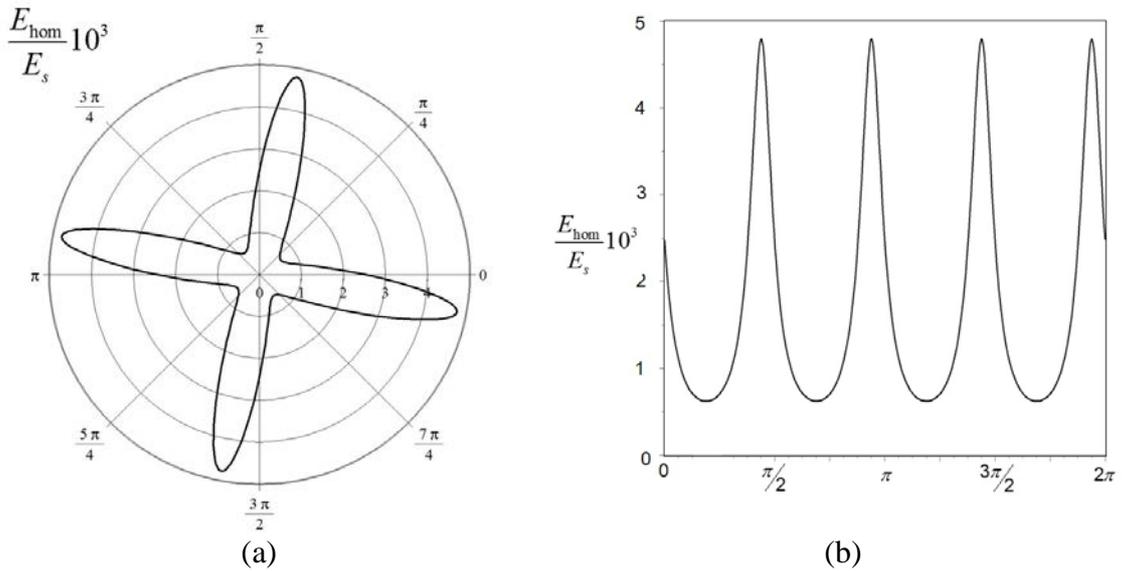

Figure 9. Polar (a) and Cartesian (b) diagrams of the ratio $E_{hom}/E_s$
($\beta = 0.38\,rad$, $\delta = 0.06$, $\rho = 0.85$).

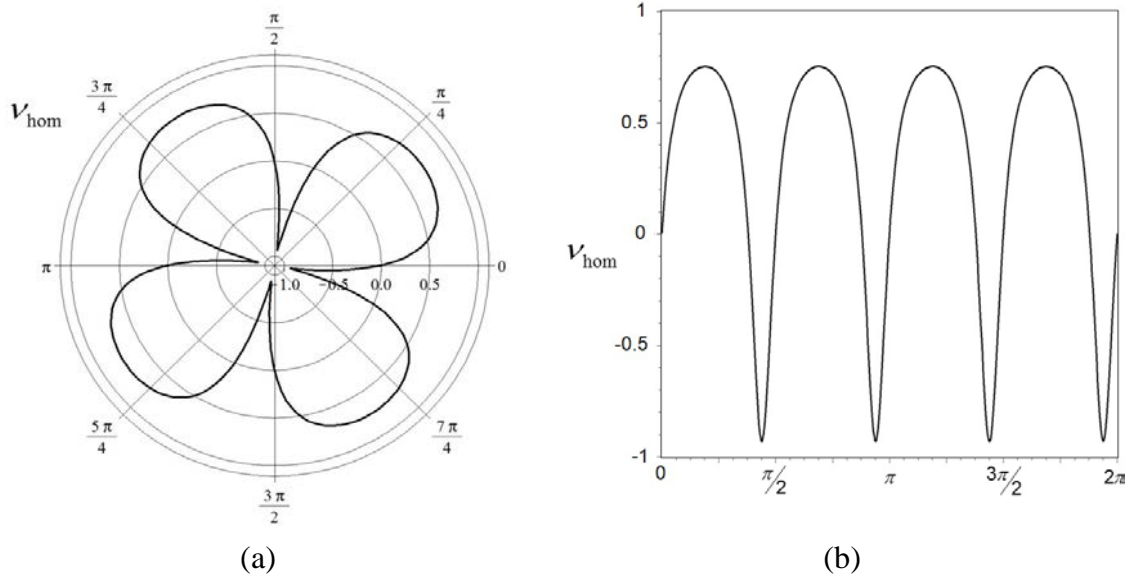

(a)             (b)

Figure 10. Polar (a) and Cartesian (b) diagrams of the overall Poisson's ratio $v_{hom}$.
($\beta = 0.38\,rad$, $\delta = 0.06$, $\rho = 0.85$).

## 3. Second gradient modelling of periodic chiral cellular solids

A higher accuracy in describing chiral periodic cellular solids may be obtained through the solid modelling of the periodic cell, conceived as a two-dimensional domain consisting of deformable portions such as the ring, the ligaments and possibly an embedded matrix internally to these. The classical homogenization of these cells has been carried out through the FE elastic analysis of the periodic cell (see for reference Alderson *et al.*, 2010a, Chen *et al.*, 2013, Dirrenberger *et al.*, 2013, Lorato *et al.*, 2010, Tee *et al.*, 2010). As the considered lattices present characteristic sizes depending on the cell geometry which are not negligible in comparison to the structural size or to the length of traveling elastic waves, it may be worthwhile to apply non-local homogenization procedures, which are able to incorporate internal lengths in the constitutive equations. Recently, Trinh *et al.*, 2012, have carried out an analysis of the non-local homogenization techniques of 2D and 3D periodic solids available in the literature. They showed that the computational homogenization techniques based on second gradient continua seem to be more reliable than those based on the micropolar continuum. In view of this, it was decided to carry out a computational homogenization of the two chiral cellular solids here considered with reference to a second displacement gradient continuum (Mindlin, 1964, Germain, 1973) according to the computational procedure proposed by Bacigalupo and Gambarotta, 2010.

At the fine scale, the micro-displacement $\mathbf{u}(\mathbf{x})$, the micro-strain tensor $\boldsymbol{\varepsilon}(\mathbf{x})=sym\nabla_x\mathbf{u}$ and the micro-stress tensor $\boldsymbol{\sigma}(\mathbf{x})$ at a point of the body located at $\mathbf{x}$ are defined. Because of the periodicity of both the geometry and the standard (Cauchy) elasticity tensor $\mathbb{C}^m$ in the lattice (see for instance figure 11.a), a periodic cell $\mathcal{A}(\mathbf{y})$ may be identified, having centre at $\mathbf{y}$, boundary $C$ and periodicity vectors $\mathbf{v}_1$, $\mathbf{v}_2$ (see figure 11.b). Accordingly, a point $\mathbf{x}$ inside the cell may be identified by the relative position vector $\mathbf{z}=\mathbf{x}-\mathbf{y}$ with respect to the centre $\mathbf{y}$.

The multi-scale kinematical model is obtained by considering the micro-displacement field in the periodic cell written in the form $\mathbf{u}(\mathbf{x})=\mathbf{u}(\mathbf{y},\mathbf{z})=\mathbf{u}^*(\mathbf{y},\mathbf{z})+\tilde{\mathbf{u}}(\mathbf{y},\mathbf{z})$, namely as the superposition of two displacement functions. The first term comes from the "quadratic ansatz" (see Trinh *et al.*, 2012)

$$\mathbf{u}^*(\mathbf{y},\mathbf{z}) = \mathbf{U}(\mathbf{y}) + \mathbf{H}(\mathbf{y})\mathbf{z} + \frac{1}{2}\boldsymbol{\kappa}(\mathbf{y}):(\mathbf{z}\otimes\mathbf{z}) \qquad (16)$$

and depends on the macro-displacement function $\mathbf{U}(\mathbf{y})$ and on the macro-strain tensors, i.e. the displacement gradient $\mathbf{H}(\mathbf{y})=\nabla_y\mathbf{U}$ and the second displacement gradient $\boldsymbol{\kappa}(\mathbf{y})=\nabla_y\otimes\mathbf{H}(\mathbf{y})$, respectively. The second contribution to the micro-displacement field represents the microstructural displacement fluctuation field $\tilde{\mathbf{u}}(\mathbf{y},\mathbf{z})$ taking into account the effects of the material heterogeneities. According to Bacigalupo and Gambarotta, 2010, this term is here assumed in the form

$$\tilde{\mathbf{u}}(\mathbf{y},\mathbf{z}) = \boldsymbol{\Theta}^1(\mathbf{z}):\left[\mathbf{H}(\mathbf{y})+\boldsymbol{\kappa}(\mathbf{y})\mathbf{z}\right] + \boldsymbol{\Theta}^2(\mathbf{z})\vdots\boldsymbol{\kappa}(\mathbf{y}), \qquad (17)$$

where $\boldsymbol{\Theta}^1(\mathbf{z})$ and $\boldsymbol{\Theta}^2(\mathbf{z})$ are a third and a fourth-order tensor, respectively.

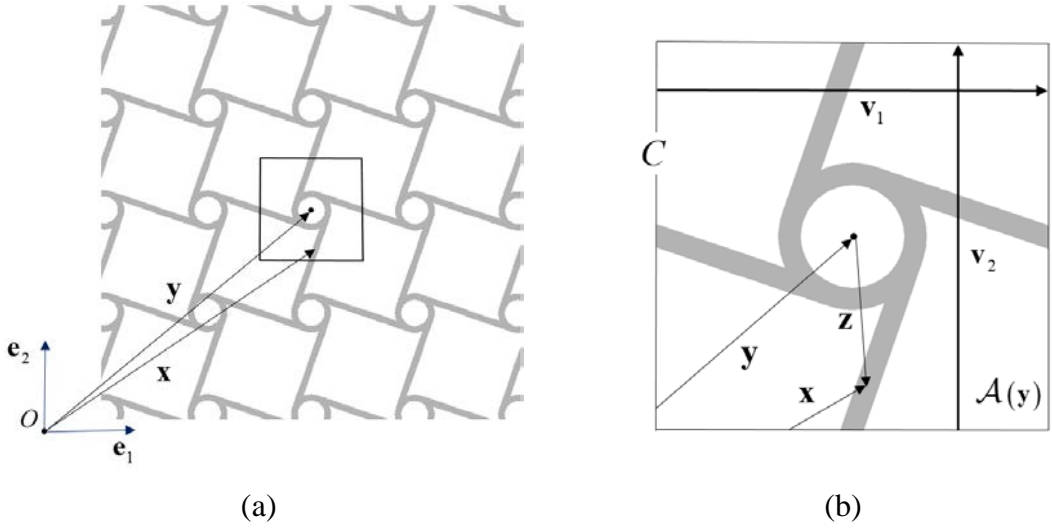

(a)  (b)

Figure 11. (a) Periodic tetrachiral composite; (b) Periodic cell and periodicity vectors for the tetrachiral composite.

As usual, the first order homogenization is carried out by considering a homogeneous macro-displacement gradient $\mathbf{H}$ in the body ($\boldsymbol{\kappa}(\mathbf{y}) = \mathbf{0}$) at the macro-scale and by imposing the classical $\mathcal{A}$-periodicity condition on the micro-displacement fluctuation $\tilde{\mathbf{u}}(\mathbf{y}, \mathbf{z}_b) = \tilde{\mathbf{u}}(\mathbf{y}, \mathbf{z}_b + \mathbf{v}_i)$, $\forall \mathbf{z}_b \in C_i$, $i = 1, n_b$, $\mathbf{z}_b$ being the local position vector at a point on the boundary $C_i$ and $n_b$ the number of periodicity vectors ($n_b = 2$ for the tetrachiral composite in figure 12). This condition implies the $\mathcal{A}$-periodicity of the function $\boldsymbol{\Theta}^1(\mathbf{z})$, i.e. $\boldsymbol{\Theta}^1(\mathbf{z}_b) = \boldsymbol{\Theta}^1(\mathbf{z}_b + \mathbf{v}_i)$, $\mathbf{z}_b \in C_i$, $i=1,n_b$ as well. Accordingly, the micro-displacement field is obtained by numerically solving a boundary value problem defined on the periodic cell with boundary conditions prescribed on the relative micro-displacement field

$$\mathbf{u}(\mathbf{y}, \mathbf{z}_b + \mathbf{v}_i) - \mathbf{u}(\mathbf{y}, \mathbf{z}_b) = \mathbf{H}\mathbf{v}_i, \quad \forall \mathbf{z}_b \in C_i, \ i = 1, n_b, \qquad (18)$$

which returns as solution the micro-displacement function $\mathbf{u}(\mathbf{y}, \mathbf{z})$ from which the perturbation function $\boldsymbol{\Theta}^1(\mathbf{z})$ is obtained.

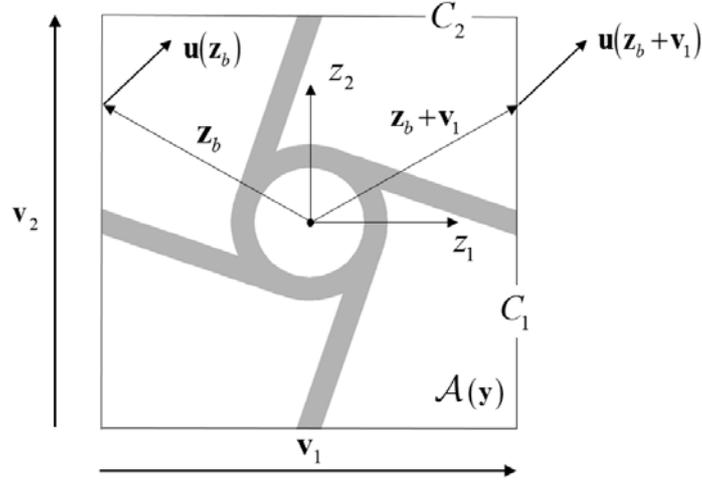

Figure 12. Displacement vectors of points at the boundary of the periodic cell for the tetrachiral composite ($n_b = 2$).

The second-order homogenization is carried out by considering a homogeneous second displacement gradient $\boldsymbol{\kappa}$ in the body at the macro-scale, so that the gradient tensor $\mathbf{H}(\mathbf{y})$ turns out to be affine. In this case, to obtain a continuous displacement fields across the periodic cell interfaces (see Bacigalupo and Gambarotta, 2010, for details) the fluctuation function $\boldsymbol{\Theta}^2(\mathbf{z})$ must be $\mathcal{A}$-periodic as well:

$$\boldsymbol{\Theta}^2(\mathbf{z}_b) = \boldsymbol{\Theta}^2(\mathbf{z}_b + \mathbf{v}_i) \qquad \mathbf{z}_b \in C_i, \ i=1,n_b. \tag{19}$$

The $2^4$ unknown functions $\theta^2_{ijkl}(\mathbf{z})$ of $\boldsymbol{\Theta}^2(\mathbf{z})$ are obtained by analysing the periodic cell with prescribed boundary conditions on the unknown micro-displacement field $\mathbf{u}^{II}(\mathbf{y},\mathbf{z})$, which take the form

$$\mathbf{u}^{II}(\mathbf{z}_b + \mathbf{v}_i) - \mathbf{u}^{II}(\mathbf{z}_b) = \mathbf{u}^*(\mathbf{z}_b + \mathbf{v}_i) - \mathbf{u}^*(\mathbf{z}_b) + \boldsymbol{\Theta}^1(\mathbf{z}_b) : \boldsymbol{\kappa}\mathbf{v}_i \qquad \mathbf{z}_b \in C_i, \ i=1,n_b, \tag{20}$$

where the macro-position $\mathbf{y}$ is omitted for simplicity. With reference to the periodic cell of the tetrachiral composite shown in figure 12, the boundary conditions referred to the vertical side $C_1$ and horizontal side $C_2$ are written in components in the following form, respectively:

$$u_i^{II}(z_1^+) - u_i^{II}(z_1^-) = u_i^*(z_1^+) - u_i^*(z_1^-) + \theta_{i11}^{1+}d_1\kappa_{111} + \theta_{i12}^{1+}d_1\kappa_{112} + \theta_{i21}^{1+}d_1\kappa_{211} + \theta_{i22}^{1+}d_1\kappa_{221},$$
$$u_i^{II}(z_2^+) - u_i^{II}(z_2^-) = u_i^*(z_2^+) - u_i^*(z_2^-) + \theta_{i11}^{1+}d_2\kappa_{112} + \theta_{i12}^{1+}d_2\kappa_{122} + \theta_{i21}^{1+}d_2\kappa_{212} + \theta_{i22}^{1+}d_2\kappa_{222},$$
(21)

being $z_i^\pm = \pm d_i/2$, $i=1,2$, $\theta_{hkl}^{1+} = \theta_{hkl}^1(z_i^+)$, $d_i = |\mathbf{v}_i|$ and $\kappa_{ijk} = \dfrac{\partial^2 U_i}{\partial y_j \partial y_k}$. The corresponding micro-displacement field $\mathbf{u}^{II}(\mathbf{y},\mathbf{z})$ is obtained by a FE analysis of the periodic cell so that the unknown functions $\theta_{iklp}^2(\mathbf{z})$ may be derived from equations

$$\kappa_{klp}\theta_{iklp}^2(\mathbf{z}) = u_i^{II}(\mathbf{z}) - u_i^*(\mathbf{z}) - \theta_{ikl}^1(\mathbf{z})\kappa_{klp}z_p, \qquad (22)$$

where the functions at the r.h.s. are known from the previous analysis.

The elastic moduli of the second-order continuum are evaluated in terms of the macro-strain vectors $\underline{E} = \{H_{11} \;\; H_{22} \;\; H_{12}+H_{21}\}^T$ and $\underline{\kappa} = \{\kappa_{111} \;\; \kappa_{222} \;\; \kappa_{122} \;\; \kappa_{211} \;\; \sqrt{2}\kappa_{121} \;\; \sqrt{2}\kappa_{212}\}^T$ represented in the Voigt notation, where the symmetry of the second displacement gradient $\kappa_{ijk} = \kappa_{ikj}$ is taken into account. The mean value of the micro-strain energy of the periodic cell $\mathcal{E}_m = \dfrac{1}{2A}\displaystyle\int_{A(\mathbf{y})} \underline{\varepsilon}^T \underline{\underline{C}}^m \underline{\varepsilon}\, da$ may be expressed in terms of the macro-strains by noting that the micro-strain field $\underline{\varepsilon} = \{\varepsilon_{11} \;\; \varepsilon_{22} \;\; 2\varepsilon_{12}\}^T$ in the heterogeneous cell may be written in the following linear form

$$\underline{\varepsilon} = \underline{\underline{B}}^E(\mathbf{z})\underline{E} + \underline{\underline{B}}^\kappa(\mathbf{z})\underline{\kappa}, \qquad (23)$$

$\underline{\underline{B}}^E(\mathbf{z})$ and $\underline{\underline{B}}^\kappa(\mathbf{z})$ being localization matrices depending on the functions $\theta_{ikl}^1(\mathbf{z})$ and $\theta_{iklp}^2(\mathbf{z})$, i.e. on the periodic microstructure of the cell. On the other side, the macro-strain energy $\mathcal{E}_M$ at a point $\mathbf{y}$ of the homogenized continuum depends on the macro-strain vectors as follows

$$\mathcal{E}_M(\underline{E},\underline{\kappa}) = \frac{1}{2}\{\underline{E}^T \;\; \underline{\kappa}^T\}\begin{bmatrix}\underline{\underline{C}} & \underline{\underline{Y}} \\ \underline{\underline{Y}}^T & \underline{\underline{S}}\end{bmatrix}\begin{Bmatrix}\underline{E} \\ \underline{\kappa}\end{Bmatrix}, \qquad (24)$$

$\underline{\underline{C}}$, $\underline{\underline{Y}}$ and $\underline{\underline{S}}$ being the sub-matrices of the second-order elastic stiffness matrix. These elastic matrices are then obtained by applying the Hill-Mandel macro-homogeneity condition $\mathcal{E}_M = \mathcal{E}_m$ and take the form

$$\underline{\underline{C}} = \frac{1}{A}\int_A \underline{\underline{B}}^{E\mathrm{T}} \underline{\underline{C}}^m \underline{\underline{B}}^E \mathrm{d}a, \quad \underline{\underline{Y}} = \frac{1}{A}\int_A \underline{\underline{B}}^{E\mathrm{T}} \underline{\underline{C}}^m \underline{\underline{B}}^\kappa \mathrm{d}a, \quad \underline{\underline{S}} = \frac{1}{A}\int_A \underline{\underline{B}}^{\kappa\mathrm{T}} \underline{\underline{C}}^m \underline{\underline{B}}^\kappa \mathrm{d}a, \quad (25)$$

with $\underline{\underline{C}}$ and $\underline{\underline{S}}$ symmetric matrices (the matrix $\underline{\underline{C}}$ is the classical elasticity matrix). In general, the overall stiffness matrix of the second-order elastic plane model is characterised by 45 elasticities (matrix $\underline{\underline{C}}$ - 6 elasticities, matrix $\underline{\underline{S}}$ - 21 elasticities, matrix $\underline{\underline{Y}}$ -18 elasticities). In case of centro-symmetric periodic cell one obtains $\underline{\underline{Y}} = \underline{\underline{0}}$.

Finally, the real stress tensor $\mathbf{T}(\mathbf{y}) = \mathbf{\Sigma}(\mathbf{y}) - Div_\mathbf{y}\mathbf{\mu}(\mathbf{y})$ in the homogenized medium is obtained from the first-order stress tensor $\mathbf{\Sigma}$ and the second-order stress tensor $\mathbf{\mu}$, represented in the Voigt notation by the vector $\underline{\Sigma} = \{\Sigma_{11} \quad \Sigma_{22} \quad \Sigma_{12}\}^\mathrm{T}$ and vector $\underline{\mu} = \{\mu_{111} \quad \mu_{222} \quad \mu_{122} \quad \mu_{211} \quad \sqrt{2}\mu_{121} \quad \sqrt{2}\mu_{212}\}^\mathrm{T}$, respectively. From the assumption of macro-strain elastic energy, the above defined stress vectors are obtained in the following form

$$\begin{aligned}\underline{\Sigma} &= \frac{\partial \mathcal{E}_M}{\partial \underline{E}} = \underline{\underline{C}}\,\underline{E} + \underline{\underline{Y}}\,\underline{\kappa}, \\ \underline{\mu} &= \frac{\partial \mathcal{E}_M}{\partial \underline{\kappa}} = \underline{\underline{Y}}^\mathrm{T}\underline{E} + \underline{\underline{S}}\,\underline{\kappa}.\end{aligned} \quad (26)$$

### 3.1 Hexachiral honeycomb

The hexachiral cell is shown in Figure 13, with $n_b = 3$ periodicity vectors. Because of the hexagonal symmetry, the classical (Cauchy) elasticities $C_{ijhk}$ are those ones of a transverse isotropic continuum, i.e. the in-plane Young's modulus $E_{\mathrm{hom}}$ and the Poisson's ratio $\nu_{\mathrm{hom}}$. Since the cell is centro-symmetric, it follows $Y_{ijhkp} = 0$. In addition, the elasticities $S_{ijhkpq}$ of the second order elasticity tensor, which depend on the chirality parameter $\beta$, must satisfy the conditions of invariance respect to the cyclic group $Z_6$, i.e. the group of rotations $\mathbf{R}_3^{\pi/3}$ of the chiral figure that possesses a 6-fold invariance. Auffray *et al.*, 2009, derived the properties of the second-order elasticities within the second-order strain-gradient theory by Mindlin (Mindlin, 1964, Mindlin and Eshel, 1968), which are discussed in Appendix. Through the relations established by Mindlin and Eschel, 1968, between the strain gradient elasticity formulation (form II) and the second displacement gradient formulation (form I) considered here, it appears that the elastic stiffness matrix in the form I

$$\underline{\underline{S}} = \begin{pmatrix} S_{111111} & S_{111222} & S_{111122} & S_{111211} & \sqrt{2}S_{111121} & \sqrt{2}S_{111212} \\ & S_{222222} & S_{222122} & S_{222211} & \sqrt{2}S_{222121} & \sqrt{2}S_{222212} \\ & & S_{122122} & S_{122211} & \sqrt{2}S_{122121} & \sqrt{2}S_{122212} \\ & & & S_{211211} & \sqrt{2}S_{211121} & \sqrt{2}S_{211212} \\ & SYM & & & 2S_{121121} & 2S_{121212} \\ & & & & & 2S_{212212} \end{pmatrix}$$

is characterized by 7 elasticities $S_{111111}$, $S_{222222}$, $S_{212212}$, $S_{211211}$, $S_{111121}$, $S_{222122}$, $S_{111212}$. In fact, the remaining elasticities depend on the previous ones in the form:

$$S_{111222} = -2S_{111121} + S_{222122}, \qquad S_{111211} = 4S_{111121} - 3S_{222122},$$
$$S_{111122} = S_{111111} - 2S_{212212} + \frac{1}{2}(S_{211211} - S_{222222}), \qquad S_{122211} = S_{222122} - 2S_{111121},$$
$$S_{222211} = 2(S_{111111} - S_{212212}) + \frac{1}{2}(S_{211211} - 3S_{222222}), \qquad S_{222121} = S_{111111} - S_{222222} + S_{111212},$$
$$S_{222212} = 3S_{111121} - 2S_{222122}, \qquad S_{122122} = S_{111111} - S_{222222} + S_{211211}, \qquad (27)$$
$$S_{122121} = -3S_{111121} + 2S_{222122}, \qquad S_{122212} = -S_{111212} + \frac{1}{2}(S_{222222} - S_{211211}),$$
$$S_{211121} = -S_{111111} - S_{111212} + \frac{1}{2}(3S_{222222} - S_{211211}), \qquad S_{211212} = -S_{111121},$$
$$S_{121121} = -S_{111111} + S_{222222} + S_{212212}, \qquad S_{121212} = S_{222122} - 2S_{111121}.$$

Finally, from equations (27) one obtains $S_{222212} = -S_{122121}$, $S_{211212} = -S_{111121}$, which are odd functions of the chirality angle $\beta$.

*3.2 Tetrachiral honeycomb*

The tetrachiral cell is shown in Figure 14, with $n_b = 2$ periodicity vectors. In this case, the classical elasticities $C_{ijhk}$ are those of the tetragonal system (4 in-plane elasticities) according to the constitutive equation (11). Because of the centro-symmetry of the cell it follows $Y_{ijhkp} = 0$, while the non-local elasticities $S_{ijhkpq}$ have to satisfy the condition of invariance with respect to the cyclic group $Z_4$, namely i.e. the group of rotations $\mathbf{R}_3^{\pi/2}$ of the chiral figure that possesses a 4-fold invariance. In this case, according to Auffray *et al.*, 2009, there are 9 elasticities $S_{111111}$, $S_{111122}$, $S_{111121}$, $S_{111212}$, $S_{222122}$, $S_{211211}$, $S_{211121}$, $S_{211212}$, $S_{212212}$, and the remaining ones are

$$\begin{aligned}
&S_{122122} = S_{211211}, \quad S_{111211} = -S_{222122}, \quad S_{222211} = S_{111122}, \\
&S_{222121} = S_{111212}, \quad S_{222212} = -S_{111121}, \quad S_{122212} = S_{211121}, \\
&S_{222222} = S_{111111}, \quad S_{122121} = -S_{211212}, \quad S_{121121} = S_{212212}, \\
&S_{111222} = S_{122211} = S_{121212} = 0.
\end{aligned} \quad (28)$$

Also in this case, the chiral geometry of the microstructure affects the non-local elasticities $S_{ijhkpq}$, and from equations (28) one obtains $S_{111211} = -S_{222122}$, $S_{222212} = -S_{111121}$, $S_{122121} = -S_{211212}$, which are odd functions of the chirality angle $\beta$.

## 4. Second gradient homogenization: numerical examples

As a first example, the hexachiral material shown in Figure 1.a is considered, whose constituents are assumed isotropic, perfectly bonded, undergoing plane stress condition. With reference to Figure 2.a, the following values for the model parameters are assumed: $R = 5$ mm, $t = 1.5$ mm, $l = 25$ mm, $E_s = 1600$ MPa and $\nu_s = 0.36$. These data (with the exception of $\nu_s$) correspond to those of the sample experienced by Alderson *et al.* (2010). The region between the ligaments is filled with a material called *m1*, while the interior of the central ring is filled with the material called *m2*. The elastic moduli of the homogenized material have been evaluated for different ratios $E_{m_1}/E_s$, $E_{m_2}/E_s$ of the filling materials in order to appreciate the effects of the stiffness mismatch between the ligaments and the filled materials on the overall local and non-local moduli. To simplify the comparisons among the results, the same Poisson's ratio $v_{m_1} = v_{m_2} = 0.36$ has been assumed for the three considered materials. In this regard, four models have been analysed, which are here refererred to as Model A, B, C and D, respectively, identified in Table 1, and which are characterized by different values of the ratios $E_{m_1}/E_s$, $E_{m_2}/E_s$. It should be noted that the Model A refers to the case in which the central disk is made from the same material of the ligaments, so that it can be regarded as rigid. This model is closer to the assumptions made in defining the micropolar model considered in Section 2 (where the ring is assumed to be rigid). Conversely, the Model B refers to a cellular solid in the absence of the filling material and corresponds to the sample experienced by Alderson *et al*., 2010. Finally, the Models C and D have been considered in order to appreciate the effect of the filling material on the overall stiffness and the related auxetic properties. This fact seems to be of interest in order to make the material stiffer, in consideration of the resultant remarkable compliance of the cellular solid if compared to that of the ligaments, a point that may constitutes a limit for the application of these composite materials.

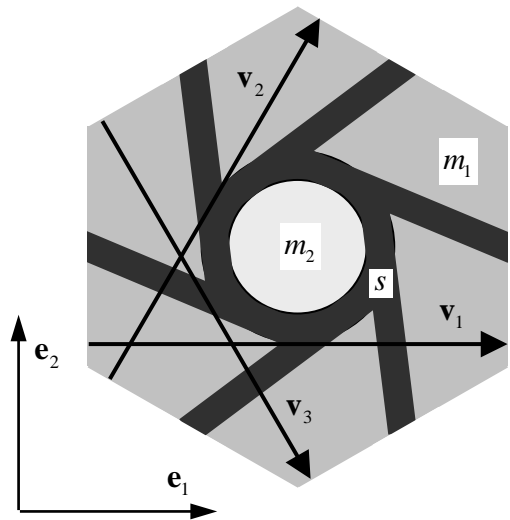

Figure 13: Hexachiral periodic cell and periodicity vectors.

Table 1. Stiffness ratios $E_{m_1}/E_s$, $E_{m_2}/E_s$ for the four considered models.

| Mod. | $E_{m_1}/E_s$ | $E_{m_2}/E_s$ |
|------|---------------|---------------|
| A    | 0             | 1             |
| B    | 0             | 0             |
| C    | $1 \cdot 10^{-2}$ | $1 \cdot 10^{-2}$ |
| D    | $1 \cdot 10^{-1}$ | $1 \cdot 10^{-1}$ |

The overall elastic moduli $C_{ijhk}$ and $S_{ijhkpq}$, evaluated through the second-order computational homogenization presented in Section 3, are given in Table 2, together with the Young's modulus $E_{\text{hom}}$ and Poisson's ratio of the homogenized material $\nu_{\text{hom}}$.

Table 2. Hexachiral cellular solid – Elastic moduli $C_{ijhk}$ (MPa), $S_{ijhkpq}$ (MPa mm$^2$).

| Mod. | $C_{1111} = C_{2222}$ | $C_{1122}$ | $C_{1212}$ | $E_{\text{hom}}$ | $\nu_{\text{hom}}$ |
|------|-----------------------|------------|------------|------------------|---------------------|
| A    | 49.24                 | −38.13     | 43.73      | 19.71            | −0.77               |
| B    | 29.95                 | −19.30     | 24.63      | 17.51            | −0.64               |
| C    | 70.02                 | −2.27      | 36.19      | 70.00            | −0.03               |
| D    | 306.15                | 100.00     | 103.29     | 273.92           | 0.33                |

| Mod. | $S_{111111}$ | $S_{111121}$ | $S_{111212}$ | $S_{222222}$ | $S_{222122}$ | $S_{211211}$ | $S_{212212}$ |
|------|--------------|--------------|--------------|--------------|--------------|--------------|--------------|
| A    | 99.73        | −39.24       | 34.56        | 132.05       | −72.86       | 86.72        | 27.37        |
| B    | 98.32        | −38.42       | 34.18        | 128.42       | −71.62       | 83.67        | 26.96        |
| C    | 351.62       | 43.71        | 42.61        | 373.67       | 32.95        | 272.44       | 174.28       |
| D    | 1823.48      | 637.58       | −51.01       | 1775.74      | 844.91       | 1805.88      | 1376.34      |

The second-order stiffness matrix for Model B is

$$\underline{\underline{S}} = \begin{pmatrix} 98.32 & 5.22 & 22.04 & 61.19 & -38.42 & 34.18 \\ & 128.42 & -71.62 & -8.06 & 4.08 & 27.99 \\ & & 53.57 & 5.22 & -27.99 & -11.80 \\ & & & 83.67 & 18.29 & 38.42 \\ & & & & 57.05 & 5.22 \\ & & & & & 26.96 \end{pmatrix} \text{MPa mm}^2, \quad (29)$$

where the elasticities depending on the sign of the chirality parameter may be identified $S_{222212} = -S_{122121} = 27.99$ MPa mm$^2$, $S_{211212} = -S_{111121} = 38.42$ MPa mm$^2$. In fact, if the material is flipped over one obtains $S_{222212} = -S_{122121} = -27.99$ MPa mm$^2$, $S_{211212} = -S_{111121} = -38.42$ MPa mm$^2$, while the other components remain unchanged. The corresponding overall elasticities $S_{ijhkpq}^{SG}$ in the strain-gradient formulation are obtained through the transformation equations by Mindlin and Eschel, 1968, and are given in Table 3 for the considered models where $S_{iiiiii} = S_{iiiiii}^{SG}$, as expected (see Appendix A).

Table 3. Hexachiral cellular solid – Strain-gradient formulation– elastic moduli $S_{ijhkpq}^{SG}$ (MPa mm$^2$).

| Mod. | $S_{111111}^{SG}$ | $S_{111222}^{SG}$ | $S_{111221}^{SG}$ | $S_{111122}^{SG}$ | $S_{222222}^{SG}$ | $S_{222221}^{SG}$ | $S_{112112}^{SG}$ |
|---|---|---|---|---|---|---|---|
| A | 99.73 | 5.45 | 46.44 | 22.67 | 132.05 | 129.08 | 247.01 |
| B | 98.32 | 5.03 | 45.95 | 22.41 | 128.42 | 127.73 | 241.86 |
| C | 351.62 | −58.45 | 130.47 | −45.24 | 373.67 | 109.26 | 944.52 |
| D | 1823.48 | −463.46 | 800.43 | −902.46 | 1775.74 | −294.78 | 7196.30 |

In the strain-gradient formulation, the second-order stiffness matrix for Model B takes the form

$$\underline{\underline{S}}^{SG} = \begin{pmatrix} 98.32 & 5.03 & 45.95 & -137.80 & 31.69 & 86.76 \\ & 128.42 & 127.73 & 15.85 & -101.00 & -10.87 \\ & & 211.76 & 5.03 & -111.90 & 101.00 \\ & & & 241.86 & -86.76 & -69.34 \\ & & & & 109.09 & 10.07 \\ & & & & & 169.29 \end{pmatrix} \text{MPa mm}^2, \quad (30)$$

where the elasticities that depend on the sign of the chirality parameter may be identified $S^{SG}_{111121} = -S^{SG}_{112122} = 86.76$ MPa mm$^2$ , $S^{SG}_{222122} = -S^{SG}_{221121} = -101.00$ MPa mm$^2$ . By comparing the elastic moduli of the Model A (filled ring) with those of Model B (hollow ring) given in Table 2, appreciable differences in the elastic constants are observed, even if those ones are contained within 13% for the Young's modulus and 20% for the Poisson's ratio. The presence of the filling material between the ligaments and in the ring, with a very low ratio $E_m/E_s = 10^{-2}$ between the elastic stiffnesses (Model C) has a significant effect on the elastic stiffness. In Model C, the Poisson's ratio is almost zero, while in Model D the ratio becomes positive and the material is no longer auxetic. Finally, we note that, unlike the elastic constants of the first order, the elastic moduli of the second-order matrix $\underline{\underline{S}}$ of Model A and B, respectively, appear to be only slightly different.

As a second example let us consider the tetrachiral material shown in Figure 3.a, whose periodic cell and the vectors of periodicity are shown in Figure 14. The geometric properties of the microstructure are the same as the previous case, as well as the elastic properties of the components.

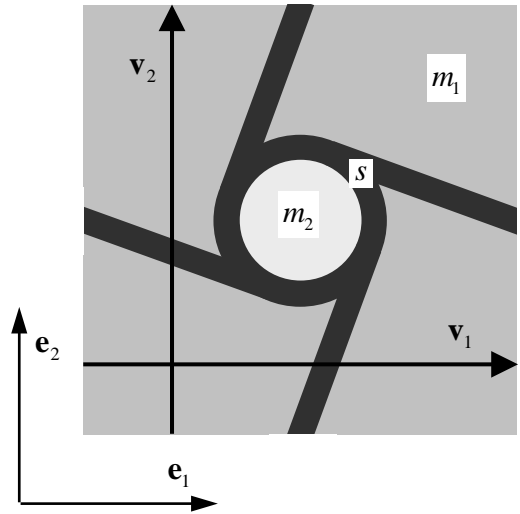

Figure 14. Thetrachiral periodic cell and periodicity vectors.

The overall elastic moduli $C_{ijhk}$ and $S_{ijhkpq}$ are given in Table 4, together with the Young's modulus $E_{\text{hom}}^0 = \Sigma_{11}/E_{11}\left(=\Sigma_{22}/E_{22}\right)$ and the Poisson's ratio $\nu_{\text{hom}}^0 = -E_{22}/E_{11}\left(=-E_{11}/E_{22}\right)$ obtained for the case of uniaxial stress applied along the direction of the unit vector $\mathbf{e}_1\,(\mathbf{e}_2)$. Among the most significant results, it is worth noting that the Poisson's ratio in Model A is zero $\nu_{\text{hom}}^0 = 0$, a result that is consistent with that obtained in Section 2.2. In agreement with the results obtained in Section 2.2, the first order elasticity tensor depends on the chirality parameter, but only the elasticities $C_{2212} = -C_{2212}$ are odd functions of the angle $\beta$.

Table 4. Tetrachiral cellular solid - Elastic moduli $C_{ijhk}$ (MPa), $S_{ijhkpq}$ (MPa mm$^2$).

| Mod. | $C_{1111} = C_{2222}$ | $C_{1122}$ | $C_{1212}$ | $C_{2212} = -C_{1112}$ | $E_{\text{hom}}^0$ | $\nu_{\text{hom}}^0$ |
|---|---|---|---|---|---|---|
| A | 47.59 | –41.21 | 5.00 | 14.36 | 6.35 | 0 |
| B | 22.95 | –17.07 | 2.78 | 7.09 | 4.60 | 0.22 |
| C | 56.91 | –4.05 | 11.51 | 9.56 | 48.66 | 0.08 |
| D | 269.01 | 81.51 | 75.44 | 12.21 | 240.92 | 0.31 |

| Mod. | $S_{111111}$ | $S_{111122}$ | $S_{111121}$ | $S_{111212}$ | $S_{222122}$ | $S_{211211}$ | $S_{211121}$ | $S_{211212}$ | $S_{212212}$ |
|---|---|---|---|---|---|---|---|---|---|
| A | 24.17 | 0 | 7.23 | 21.94 | –35.74 | 58.98 | 11.33 | 36.39 | 25.00 |
| B | 23.24 | 0 | 6.91 | 21.30 | –34.72 | 57.09 | 11.02 | 35.23 | 24.14 |
| C | 134.91 | –7.62 | 88.08 | 21.98 | –33.89 | 141.38 | –13.77 | 50.79 | 252.13 |
| D | 143.73 | –13.95 | 206.25 | 14.32 | –49.01 | 225.42 | –217.13 | 22.87 | 2387.88 |

A more comprehensive description of the in-plane elastic behavior of the tetrachiral cellular solid is provided by the Cartesian and polar diagrams of Figure 15 and 16 in which is shown the Young's modulus $E_{\text{hom}}(\theta)$ obtained from equation (14) and the Poisson's ratio $\nu_{\text{hom}}(\theta)$ obtained from equation (15), respectively. From these diagrams, and for all the considered models, one can observe a remarkable dependence of the modulus of elasticity on the orientation of the applied uniaxial tension. Moreover, a remarkable effect of the filling material is confirmed, including the case of Model C in which such material is softer. In addition, it is worth noting that in Models A, B and C, the Poisson's ratio varies significantly with the polar angle $\theta$, and is bounded from above $\nu_{\text{hom}} \leq 0.9$ and from below $\nu_{\text{hom}} \geq -0.9$. From the diagrams of Figure 16, it appears that the elastic behavior is auxetic for a narrow range of orientations. Finally, the filler material between the ligaments and inside the ring, even for vanishing values of the ratio $E_m/E_s = 10^{-2}$, has the effect of dramatically reducing the auxetic behavior, up to the case of the Model D that is not auxetic for any direction of the applied uniaxial stress. Within the framework of a classical homogenization, it results that the tetrachiral cellular solid, unlike the hexachiral one that is isotropic, exhibits mechanical performance with strong directional properties. In fact, from the diagrams of Figures 15 and 16 one may observe that a small change of the direction of the applied stress has remarkable both qualitative and quantitative effects on the response.

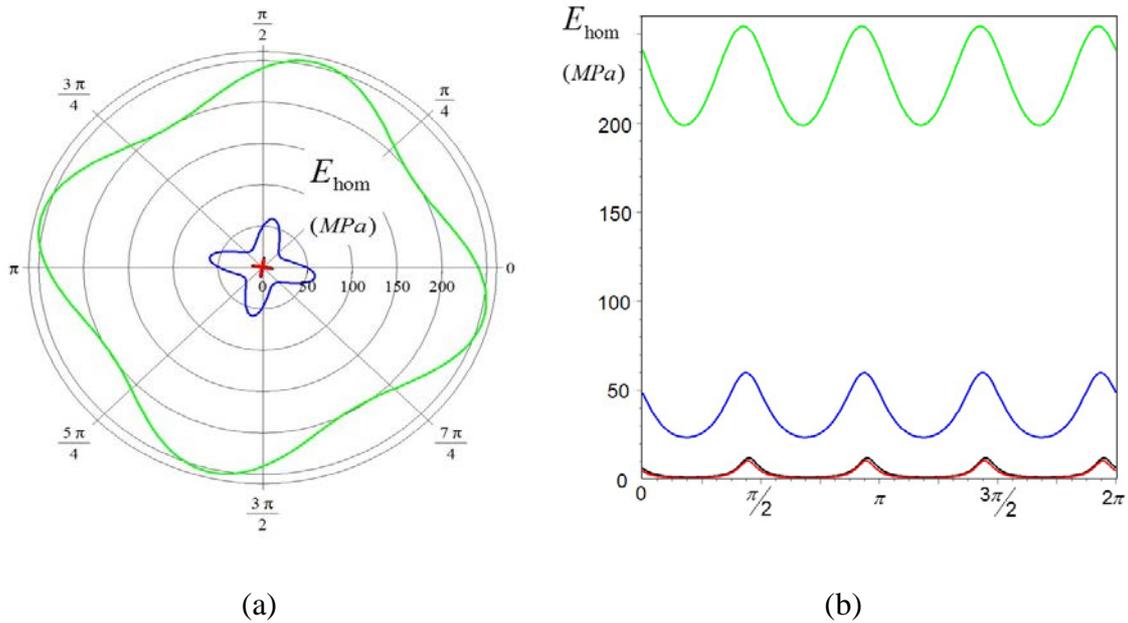

(a)                  (b)

Figure 15. Polar (a) and Cartesian (b) diagrams of the ratio $E_{\text{hom}}/E_s$ for Model A (black), B (red), C (blue), D (green).

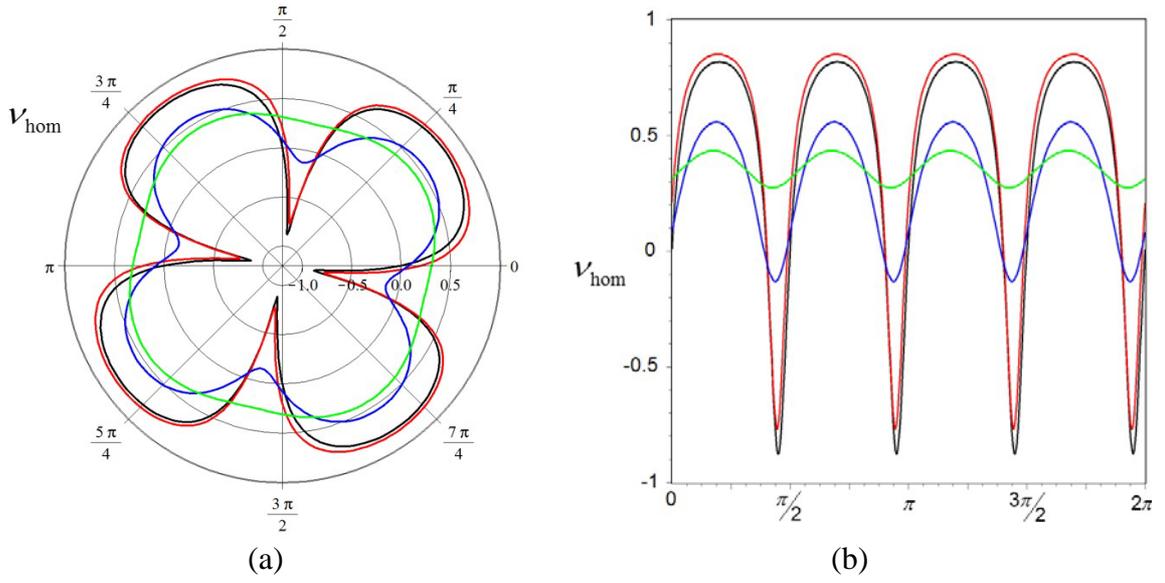

(a)                                    (b)

Figure 16. Polar (a) and Cartesian (b) diagrams of the overall Poisson's ratio $\nu_{hom}$ for Model A (black), B (red), C (blue), D (green).

The second-order stiffness matrix for Model B is

$$\underline{\underline{S}} = \begin{pmatrix} 23.24 & 0 & 0 & 34.72 & 6.91 & 21.30 \\ & 23.24 & -34.72 & 0 & 21.30 & -6.91 \\ & & 57.09 & 0 & -35.23 & 11.02 \\ & & & 57.09 & 11.02 & 35.23 \\ & & & & 24.14 & 0 \\ & & & & & 24.14 \end{pmatrix} \text{MPa mm}^2, \quad (31)$$

where the elasticities depending on the sign of the chirality parameter may be identified $S_{111211} = -S_{222122} = 34.72$ MPa mm$^2$, $S_{222212} = -S_{111121} = -6.91$ MPa mm$^2$, $S_{122121} = -S_{211212} = -35.23$ MPa mm$^2$. In fact, if the cellular material is flipped over one obtains $S_{111211} = -S_{222122} = -34.72$ MPa mm$^2$, $S_{222212} = -S_{111121} = 6.91$ MPa mm$^2$, $S_{122121} = -S_{211212} = 35.23$ MPa mm$^2$, while the other components remain unchanged. The corresponding elasticities $S^{SG}_{ijhkpq}$ in the strain-gradient formulation are given in Table 5 (see Appendix).

Table 5. Tetrachiral cellular solid - Strain-gradient formulation– elastic moduli $S^{SG}_{ijhkpq}$ $(\text{MPa mm}^2)$.

| Mod. | $S^{SG}_{111111}$ | $S^{SG}_{111221}$ | $S^{SG}_{111112}$ | $S^{SG}_{111122}$ | $S^{SG}_{111121}$ | $S^{SG}_{221221}$ | $S^{SG}_{221122}$ | $S^{SG}_{221121}$ | $S^{SG}_{122122}$ |
|---|---|---|---|---|---|---|---|---|---|
| A | 24.17 | 43.85 | −21.27 | 0 | 35.74 | 113.64 | −36.32 | 72.77 | 58.98 |
| B | 23.24 | 42.59 | −20.90 | 0 | 34.72 | 109.60 | −35.06 | 70.47 | 57.09 |
| C | 134.91 | 51.58 | 142.28 | −7.62 | 33.89 | 1205.01 | −168.93 | 101.58 | 141.38 |
| D | 143.73 | 42.60 | 363.50 | −13.95 | 49.01 | 10645.46 | −659.69 | 45.75 | 225.42 |

The strain-gradient stiffness matrix for Model B is

$$\underline{\underline{S}}^{SG} = \begin{pmatrix} 23.24 & 0 & 42.59 & -20.90 & 0 & 34.72 \\ & 23.24 & 20.90 & 42.59 & -34.72 & 0 \\ & & 109.60 & 0 & -35.06 & 70.47 \\ & & & 109.60 & -70.47 & -35.06 \\ & & & & 57.09 & 0 \\ & & & & & 57.09 \end{pmatrix} \text{MPa mm}^2, \qquad (32)$$

where the elasticities depending on the sign of the chirality parameter $\beta$ are $S^{SG}_{222221} = -S^{SG}_{111112} = 20.90$ MPa mm$^2$, $S^{SG}_{222122} = -S^{SG}_{111121} = -34.72$ MPa mm$^2$, $S^{SG}_{112122} = -S^{SG}_{221121} = -70.47$ MPa mm$^2$, which change sign if the lattice is turned upside down, in agreement with Auffray *et al.*, 2010. By comparing the diagrams of figures 15.b and 16.b, one can observe a small difference between Model A and Model B, as already noted for the hexachiral model. Moreover, as already noted in Section 2.2, the tetrachiral geometry causes the rotation of the directions of maximum stiffness and of auxetic behavior ($\nu_{\text{hom}}(\theta) \leq 0$), while for $\beta = 0$, the directions of maximum stiffness coincide with the grid axes.

Finally, it is worth noting that in tetrachiral model, the sign of the parameter of chirality (left-handed or right-handed lattice) affects both the first-order elastic moduli ($C_{2212} = -C_{1112}$) and the second-order elastic moduli ($S_{111211} = -S_{222122}$, $S_{222212} = -S_{111121}$, $S_{122121} = -S_{211212}$).

## 5. Simulation of experimental results

Alderson *et al.*, 2010, carried out uniaxial tests on samples manifactured using selective laser sintering rapid prototyping of nylon powder (Duraform). Both the hexachiral and the tetrachiral honeycombs were made having geometrical parameters R=5mm, $l = 25$ mm, $t = 1.5mm$ and $r = 4.25mm$, already considered in Section 4. Moreover, testing of the materials forming the ligaments gave a Young's modulus $E_s = 1600$ MPa. The samples were tested up to 1% or 2% applied compressive strain along the direction of the line connecting the centres of the rings. The experimental results by Alderson *et al.*, 2010, concern the overall Young's modulus and Poisson's ratio and are given in Table 6 for the hexachiral honeycomb and in Table 7 for the tetrachiral model (here referred to $E_{hom}^0$ and $\nu_{hom}^0$). Moreover, the experiments were complemented by a numerical simulation carried out through a FE homogenization (see Alderson *et al.*, 2010), whose numerical results are given in both Table 6 and 7. These experimental results have been here simulated through the micropolar and the computational homogenization techniques presented in Section 2 and 3, respectively. In the latter case, the Poisson ratio $\nu = 0.36$ has been considered for the ligaments as representative of the material employed in the experiments. In the case of micropolar homogenization, where the ligaments are modeled as beams, the connection between the ligament and the ring, which constitutes an uncertain parameter of the model, has been represented considering three distinct values of such parameter $\rho = 0.75, 0.8, 1.0$. Furthermore, with reference to the computational homogenization, the comparisons are performed by considering both the case of infilled ring (called Model A and which simulates the conditions of the beam-lattice with rigid rings), and the case of empty ring (called Model B) representing the actual geometry of the samples of Alderson *et al.*, 2010.

From the comparison between the first-order elastic moduli of the hexachiral material listed in Table 6, one can observe some differences between the experimental results and the theoretical ones. By this comparison it is observed that the beam-lattice model provides a good simulation of the overall Young's modulus for $\rho = 0.8$. However, varying this free parameter, remarkable changes in the overall modulus are obtained, which show the difficulties in the calibration of this model parameter in the micropolar homogenization. Conversely, the values of the Young modulus obtained by the computational homogenization of cells corresponding the actual geometry both by Alderson *et al.*, 2010, and by the present analysis turn out to be rather scattered. On the other hand, it is worth noting that the Young's modulus obtained from model A approximates the experimental results with an error of 1.2%. It should also be pointed out that the difference between the Young's modulus by the Model A and by the beam-lattice model for $\rho = 0.8$ is contained within 1%, while the Model B, which is the one having a more realistic geometry, it is more deformable. In any case, the two-dimensional models seem to be less stiff ( for $\rho \leq 0.8$) than those based on the beam-lattice model. In addition, the Poisson's ratio obtained through the various homogenization techniques is in good agreement with the experimental result. In the micropolar model, the influence of the parameter $\rho$ on the Poisson's ratio is confirmed, while it is not easy to explain the difference between the experimental data and numerical simulation with model B, which faithfully represents the geometry of the tested samples. Finally, it is noted that the Poisson's ratio obtained by model A is lower than the one by model B; a result that is in agreement with that of Spadoni and Ruzzene, Figures 7, 2012, by comparing the results provided by the micropolar model having rigid rings with those obtained considering the ring compliance.

Table 6. Hexachiral cellular solid – Classical elastic moduli $C_{ijhk}$ (MPa), Young's modulus $E_{\text{hom}}$ (MPa) and Poisson's ratio $\nu_{\text{hom}}$.

|  | $C_{1111} = C_{2222}$ | $C_{1122}$ | $C_{1212}$ | $E_{\text{hom}}$ | $\nu_{\text{hom}}$ |
|---|---|---|---|---|---|
| *Experiment Alderson et al. (2010)* |  |  |  | 19.46 | -0.81 |
| *FEM Alderson et al. (2010)* |  |  |  | 15.49 | -0.77 |
| *Micropolar hom. – $\rho = 1.0$* | 43.84 | -39.60 | 83.43 | 8.07 | -0.90 |
| *Micropolar hom. – $\rho = 0.8$* | 57.63 | -46.87 | 104.50 | 19.51 | -0.81 |
| *Micropolar hom. – $\rho = 0.75$* | 62.26 | -49.29 | 111.56 | 23.24 | -0.79 |
| *Computational hom. (Model A)* | 49.24 | -38.13 | 43.73 | 19.71 | -0.77 |
| *Computational hom. (Model B)* | 29.95 | -19.30 | 24.63 | 17.51 | -0.64 |

The elastic moduli, the Young's modulus $E_{\text{hom}}^0$ and the Poisson's ratio $\nu_{\text{hom}}^0$ for the tetrachiral model obtained from the experiments and from the numerical simulations are given in Table 7. From these moduli, the functions $E_{\text{hom}}(\theta)$ and $\nu_{\text{hom}}(\theta)$ have been derived by the equations (14) and (15), and shown in the diagrams of Figures 17 and 18 for the beam-lattice model ($\rho = 1.$) and for the Model A. This choice is because these two models provide the best simulation of the experimental values of Young's modulus. As already noted in Sections 2.2 and 4, a remarkable sensitivity of the elastic moduli to orientation changes of the applied uniaxial stress is found for both the models. This sensitivity of the elastic response to the direction θ of the applied stress may be considered as the primary cause of the difference between the Young's modulus $E_{\text{hom}}^0$ obtained by the theoretical models and the experimental result.

Table 7. Tetrachiral cellular solid – Classical elastic moduli $C_{ijhk}$ (MPa), Young's modulus $E_{\text{hom}}$ (MPa) and Poisson's ratio $\nu_{\text{hom}}$.

|  | $C_{1111} = C_{2222}$ | $C_{1122}$ | $C_{1212}$ | $C_{2212} = -C_{1112}$ | $E_{\text{hom}}^0$ | $\nu_{\text{hom}}^0$ |
|---|---|---|---|---|---|---|
| *Experiment Alderson et al. (2010)* |  |  |  |  | 7.08 | -0.26 |
| *FEM Alderson et al. (2010)* |  |  |  |  | 12.01 | -0.83 |
| *Micropolar hom. – $\rho = 1.0$* | 52.63 | -40.88 | 1.41 | 7.59 | 11.75 | 0 |
| *Micropolar hom. – $\rho = 0.8$* | 69.20 | -47.70 | 1.88 | 9.47 | 21.50 | 0 |
| *Micropolar hom. – $\rho = 0.75$* | 75.09 | -49.60 | 2.05 | 10.10 | 25.48 | 0 |
| *Computational hom. (Model A)* | 47.59 | -41.21 | 5.00 | 14.36 | 6.35 | 0 |
| *Computational hom. (Model B)* | 22.95 | -17.07 | 2.78 | 7.09 | 4.6 | 0.22 |

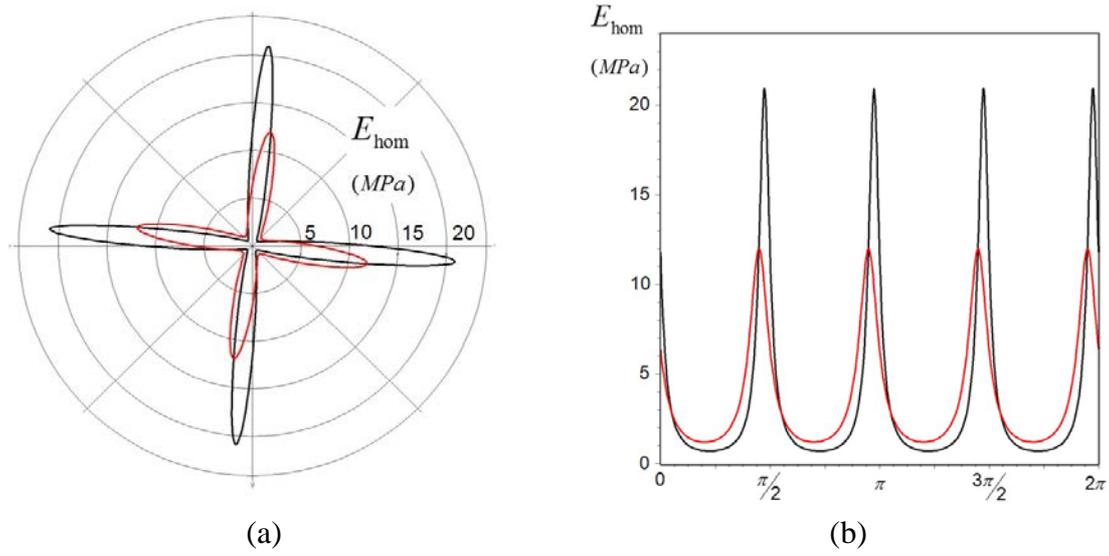

(a)          (b)

Figure 17. Comparison between the Young's modulus from the micropolar model ($\rho = 1$. - black line) and the Model A (red line).

The comparison among the values of the Young's modulus $E_{\text{hom}}^{0}$ in Table 7 shows that the Models A and B are much less stiff than the beam-lattice model, which present a remarkable sensitivity on parameter $\rho$. Nevertheless, the most relevant result is that both the beam-lattice model and the Model A (which reproduces conditions similar to the beam-lattice model) are characterized by a vanishing Poisson's ratio $\nu_{\text{hom}}^{0} = 0$, in contrast to the experimental result $\nu_{\text{hom}}^{0} = -0.83$ by Alderson *et al.*, 2010. Then, if considering the Model B, i.e. the model that reproduces the conditions of the sample experienced, a positive Poisson's ratio is obtained. These considerations highlight the difficulty in testing samples from tetrachiral cellular solids, not only for their sensitivity to the direction of the applied stress, but also because of the shear-extension coupling resulting by the non-vanishing elastic moduli $C_{2212} = -C_{1112}$, which depend on the parameter of chirality and on its sign.

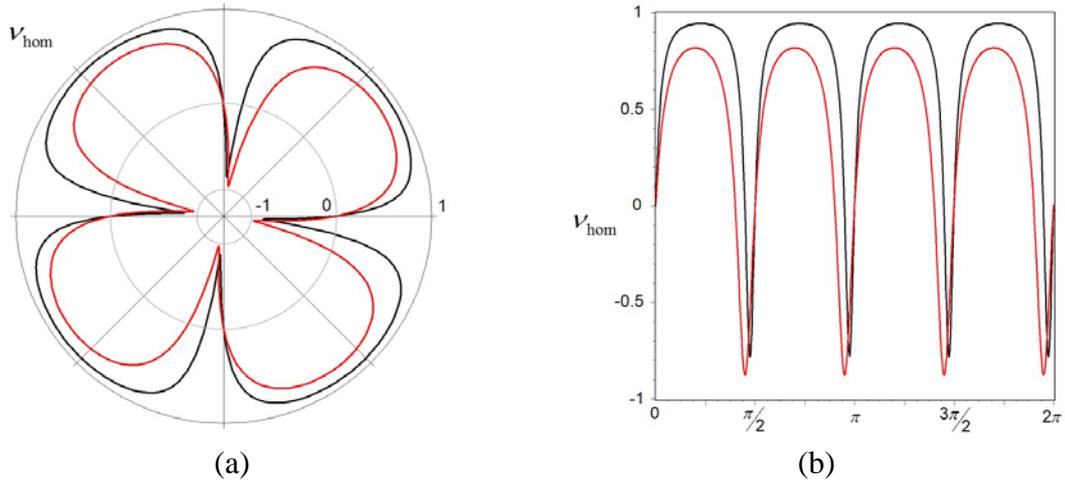

|            (a)             |            (b)             |

Figure 18. Comparison between the Poisson's ratio from the micropolar model ( $\rho = 1.$ - black line) and the Model A (red line).

## 6. Conclusions

The non-local homogenization of cellular solids having periodic hexachiral and tetrachiral microstructure has been dealt with two different techniques. The first approach, which derives from the papers of Spadoni and Ruzzene, 2012, and Liu *et al.*, 2012, is based on the representation of the cellular solid as a beam-lattice to be homogenized as a micropolar continuum. In this case, an additional parameter is introduced in the model, which describes the constraints prescribed at the ends of the ligaments to take into account of the effective connection of these to the rings, thus defining the deformable portion of the ligament. The elastic moduli of the micropolar continuum equivalent to the chiral cellular solids have been obtained. Hence, only referring to symmetric deformations, also the classical elasticities have been derived. The second approach allows to consider periodic cells having a 2-D extension and consists of a second displacement gradient computational homogenization developed by the Authors (Bacigalupo and Gambarotta, 2010). Furthermore, since both the hexachiral and the tetrachiral cells are centro-symmetric, no elastic coupling occurs in the overall constitutive equations.

The elastic moduli obtained by the micropolar homogenization are expressed in analytical form from which it appears explicitly their dependence on the parameter of chirality, which is the angle of inclination of the ligaments with respect to the grid of lines connecting the centers of the rings. For hexachiral cells, the solution of Liu *et al.*, 2012, is found, with the elastic coupling modulus between the normal and the asymmetric strains. Moreover, the classical elastic moduli obtained in case of symmetric deformation turn out to depend on the chirality parameter, but because of the transverse isotropy of the material, they do not distinguish between the left-hand and the right-hand microstructure. Also the tetrachiral cell exhibits a coupling between the normal and the asymmetric strains in the micropolar continuum through an elastic modulus that is an odd function of the parameter of chirality. Unlike the hexagonal honeycomb, the classical constitutive equations of the tetragonal honeycomb are characterized by the coupling between the normal and shear strains through an elastic modulus that is an odd function of the parameter of chirality.

The properties of the equivalent micropolar continuum here mentioned are qualitatively detected also in the equivalent second-gradient continuum. Moreover, for both the hexachiral and the tetrachiral cellular material, the second-order elastic moduli obtained through the homogenization technique here considered satisfy the $Z_6 -$ and the $Z_4 -$ invariance properties, respectively, defined by Auffray *et al.*, 2009, 2010, 2013. The classical elastic moduli are directly obtained by a second-order homogenization approach. The overall elastic moduli of the hexachiral cell are those of the transversely isotropic material and are even functions of the angle of chirality. Conversely, the elastic constants of the tetrachiral cell that couple the normal strains to the shearing strains turn out to be odd functions of the parameter of chirality. The effects of filling with elastic material between the ligaments and inside the circular rings has been analyzed through the second-order homogenization. This investigation, which is justified by the need of increasing the overall stiffness of the cellular material, showed that it is sufficient a very soft filling material to get significant increases in the Poisson's ratio, until to lose the auxetic property of the material. A similar effect is obtained for the tetrachiral material, although in this case the overall behavior is a bit more complex because of the different material symmetry (tetragonal symmetry).

The simulation of the experimental results of Alderson *et al.*, 2010, on hexachiral cell samples has shown the Young's modulus here obtained for Model B (representative of the samples) to be more reliable than that obtained by Alderson *et al.*, 2010, although both models turn out to be more deformable with respect to the behavior observed in the experiments. In contrast, the beam-lattice model appears to be generally stiffer and the Young's modulus and Poisson's ratio are greatly influenced by the parameter $\rho$, which is difficult to estimate. However, the best simulation of the experimental data is obtained for $\rho = 0.8$. The simulation of tetrachiral samples has shown a considerable variability of the Young's modulus and of the Poisson's ratio with the direction of the applied uniaxial stress. This circumstance may explain the remarkable differences between the experimental results and the numerical simulations. In this case, the Young's modulus estimated by the computational homogenisation provides a better simulation of the experimental results in comparison to the beam-lattice model, which is appears to be much stiffer. The estimates of Poisson's ratio provided by the micropolar and the second-order homogenization are consistent with each other, but disagree qualitatively with the experimental results and the numerical simulations by Alderson et al., 2010. These differences may be attributed to the variability of the response with the direction of the applied uniaxial stress. In any case, the results of the simulations suggest the need to acquire a further experimental knowledge of both the analysed chiral cellular solids.

**Acknowledgements**

The authors acknowledge financial support of the (MURST) Italian Department for University and Scientific and Technological Research in the framework of the research MIUR Prin09 project XWLFKW, *Multi-scale modelling of materials and structures*, coordinated by prof. A. Corigliano.

**Appendix**

According to the strain-gradient formulation (form II) by Mindlin, 1964, the second-order constitutive equation $\underline{\mu}^{SG} = \underline{\underline{S}}^{SG} \underline{\kappa}^{SG}$ of centro-symmetric materials, that relates the strain-gradient vector $\underline{\kappa}^{SG} = \left\{ \kappa_{111}^{SG} \quad \kappa_{222}^{SG} \quad \kappa_{221}^{SG} \quad \kappa_{112}^{SG} \quad \sqrt{2}\kappa_{122}^{SG} \quad \sqrt{2}\kappa_{121}^{SG} \right\}^T$, being $\kappa_{ijk}^{SG} = \frac{1}{2}(U_{i,j} + U_{j,i})_{,k}$, to the corresponding higher-order stress vector $\underline{\mu}^{SG} = \left\{ \mu_{111}^{SG} \quad \mu_{222}^{SG} \quad \mu_{221}^{SG} \quad \mu_{112}^{SG} \quad \sqrt{2}\mu_{122}^{SG} \quad \sqrt{2}\mu_{121}^{SG} \right\}^T$, depends on the strain-gradient stiffness matrix

$$\underline{\underline{S}}^{SG} = \begin{pmatrix} S_{111111}^{SG} & S_{111222}^{SG} & S_{111221}^{SG} & S_{111112}^{SG} & \sqrt{2}S_{111122}^{SG} & \sqrt{2}S_{111121}^{SG} \\ & S_{222222}^{SG} & S_{222221}^{SG} & S_{222112}^{SG} & \sqrt{2}S_{222122}^{SG} & \sqrt{2}S_{222121}^{SG} \\ & & S_{221221}^{SG} & S_{221112}^{SG} & \sqrt{2}S_{221122}^{SG} & \sqrt{2}S_{221121}^{SG} \\ & & & S_{112112}^{SG} & \sqrt{2}S_{112122}^{SG} & \sqrt{2}S_{112121}^{SG} \\ & SYM & & & 2S_{122122}^{SG} & 2S_{122121}^{SG} \\ & & & & & 2S_{121121}^{SG} \end{pmatrix}.$$

Auffray *et al.*, 2009, have shown that in case of plane strain-gradient elasticity, 7 elasticities are needed to describe the second-order elasticity of a two-dimensional material with $Z_6$-invariance, namely the group of the six rotations $\mathbf{R}_3^{n\pi/3}$ ($n=1,6$) of the hexachiral honeycomb. Given the elasticities $S^{SG}_{111111}$, $S^{SG}_{111222}$, $S^{SG}_{111222}$, $S^{SG}_{111221}$, $S^{SG}_{222221}$, $S^{SG}_{111122}$, $S^{SG}_{112112}$, the remaining ones (14) are obtained as follows:

$$S^{SG}_{111112} = -2S^{SG}_{111222} - S^{SG}_{222221}, \qquad S^{SG}_{111121} = \frac{1}{2}\left(-S^{SG}_{111222} + S^{SG}_{222221}\right),$$

$$S^{SG}_{111121} = S^{SG}_{111111} - S^{SG}_{222222} + S^{SG}_{111221}, \qquad S^{SG}_{222122} = -\frac{1}{2}\left(3S^{SG}_{111222} + S^{SG}_{222221}\right),$$

$$S^{SG}_{222121} = S^{SG}_{111111} - S^{SG}_{222222} + S^{SG}_{111122}, \qquad S^{SG}_{221221} = S^{SG}_{111111} - S^{SG}_{222222} + S^{SG}_{112112},$$

$$S^{SG}_{221112} = S^{SG}_{111222}, \qquad S^{SG}_{221122} = -S^{SG}_{111122} + \frac{1}{2}\left(S^{SG}_{222222} - S^{SG}_{112112}\right),$$

$$S^{SG}_{221121} = \frac{1}{2}\left(3S^{SG}_{111222} + S^{SG}_{222221}\right), \qquad S^{SG}_{112122} = \frac{1}{2}\left(S^{SG}_{111222} - S^{SG}_{222221}\right), \qquad (33)$$

$$S^{SG}_{112121} = -S^{SG}_{111111} - S^{SG}_{111122} + \frac{1}{2}\left(3S^{SG}_{222222} - S^{SG}_{112112}\right), \qquad S^{SG}_{122121} = S^{SG}_{111222},$$

$$S^{SG}_{122122} = \frac{1}{2}\left[S^{SG}_{111111} - S^{SG}_{111221} - \frac{1}{2}\left(S^{SG}_{222222} - S^{SG}_{112112}\right)\right],$$

$$S^{SG}_{121121} = -\frac{1}{2}\left[S^{SG}_{111111} + S^{SG}_{111221} - \frac{1}{2}\left(3S^{SG}_{222222} + S^{SG}_{112112}\right)\right].$$

Furthermore, they have shown how the chiral geometry entails the following relations $S^{SG}_{111121} = -S^{SG}_{112122}$, $S^{SG}_{222122} = -S^{SG}_{221121}$, the sign of which depends on the sign of the parameter of chirality $\beta$.

In the case of tetrachiral honeycomb, the non-local elasticities have to satisfy the conditions of invariance with respect to the group $Z_4$, i.e. the group of the four rotations $\mathbf{R}_3^{n\pi/2}$ ($n=1,4$) and 9 elasticities are needed. Given the elasticities $S^{SG}_{111111}$, $S^{SG}_{221221}$, $S^{SG}_{122122}$, $S^{SG}_{111221}$, $S^{SG}_{111112}$, $S^{SG}_{111122}$, $S^{SG}_{111121}$, $S^{SG}_{221122}$, $S^{SG}_{221121}$, the remaining ones (12) are obtained as follows:

$$S^{SG}_{222222} = S^{SG}_{111111}, \qquad S^{SG}_{222221} = -S^{SG}_{111112}, \qquad S^{SG}_{222112} = S^{SG}_{111221},$$
$$S^{SG}_{222122} = -S^{SG}_{111121}, \qquad S^{SG}_{222121} = S^{SG}_{111122}, \qquad S^{SG}_{112112} = S^{SG}_{221221}, \qquad (34)$$
$$S^{SG}_{112122} = -S^{SG}_{221121}, \qquad S^{SG}_{112121} = S^{SG}_{221122}, \qquad S^{SG}_{121121} = S^{SG}_{122122},$$
$$S^{SG}_{111222} = S^{SG}_{221112} = S^{SG}_{122121} = 0.$$

Also in this case, the following relations hold $S^{SG}_{222221} = -S^{SG}_{111112}$, $S^{SG}_{222122} = -S^{SG}_{111121}$, $S^{SG}_{112122} = -S^{SG}_{221121}$, the sign of which depends on the sign of the parameter of chirality $\beta$.